\documentclass[twocolumn,showpacs,pre,floatfix,superscriptaddress]{revtex4}
\usepackage{color} 
\usepackage{tabularx} 
\usepackage{epsfig}
\usepackage{amsmath} 
\usepackage{amssymb} 
\usepackage{graphicx}
\usepackage{wasysym}

\newcommand{\lglo}{\langle\!\langle}
\newcommand{\lgle}{\langle\!\langle\!\langle}
\newcommand{\blglo}{\left\langle\!\!\!\left\langle}
\newcommand{\blgle}{\left\langle\!\!\!\left\langle\!\!\!\left\langle}

\newcommand{\rglo}{\rangle\!\rangle}
\newcommand{\rgle}{\rangle\!\rangle\!\rangle}
\newcommand{\brglo}{\right\rangle\!\!\!\right\rangle}
\newcommand{\brgle}{\right\rangle\!\!\!\right\rangle\!\!\!\right\rangle}

\begin{document}

\title{Boolean decision problems with competing interactions on
scale-free networks:\\ Critical thermodynamics}

\author{Helmut G.~Katzgraber}
\affiliation {Department of Physics and Astronomy, Texas A\&M University,
College Station, Texas 77843-4242, USA}
\affiliation {Theoretische Physik, ETH Zurich, CH-8093 Zurich, Switzerland}

\author{Katharina Janzen}
\affiliation{Institut f\"{u}r Mathematische Physik, TU Braunschweig, D-38106
Braunschweig, Germany\footnote{present address}}
\affiliation{Institut f\"{u}r Physik, Carl-von-Ossietzky-Universit{\"a}t,
D-26111 Oldenburg, Germany}

\author{Creighton K.~Thomas}
\affiliation {Department of Physics and Astronomy, Texas A\&M University,
College Station, Texas 77843-4242, USA}
\affiliation {Department of Materials Science and Engineering, Northwestern
University, Evanston, Illinois 60208-3108, USA$^\ast$}

\date{\today}

\begin{abstract}

We study the critical behavior of Boolean variables on scale-free
networks with competing interactions (Ising spin glasses). Our
analytical results for the disorder--network-decay-exponent phase
diagram are verified using Monte Carlo simulations.  When
the probability of positive (ferromagnetic) and negative
(antiferromagnetic) interactions is the same, the system undergoes a
finite-temperature spin-glass transition if the exponent that describes
the decay of the interaction degree in the scale-free graph is strictly
larger than 3. However, when the exponent is equal to or less than 3, a
spin-glass phase is stable for all temperatures. The robustness of both
the ferromagnetic and spin-glass phases suggests that Boolean decision
problems on scale-free networks are quite stable to local perturbations.
Finally, we show that for a given decay exponent spin glasses on
scale-free networks seem to obey universality. Furthermore,
when the decay exponent of the interaction degree is larger than 4 in
the spin-glass sector, the universality class is the same as for the
mean-field Sherrington-Kirkpatrick Ising spin glass.

\end{abstract}

\pacs{75.50.Lk, 75.40.Mg, 05.50.+q, 64.60.-i}

\maketitle

\section{Introduction and Motivation}
\label{sec:introduction}

\begin{figure}[!tb]
\includegraphics[width=\columnwidth]{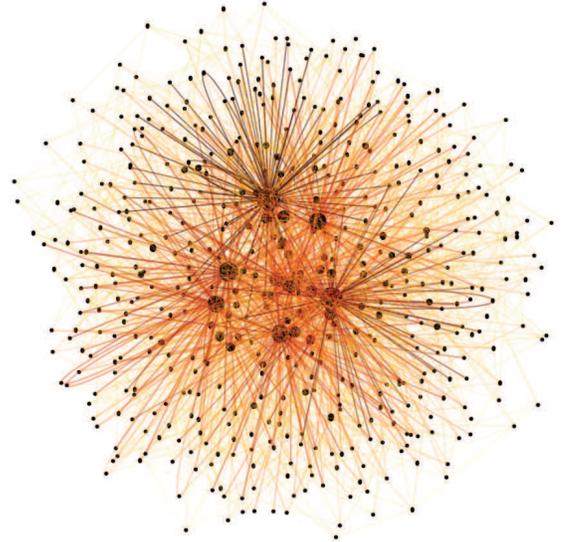}
\caption{(Color online)
Typical network simulated. The connectivity matrix is selected
according to Eq.~(\ref{eq:p_k}) (i.e., the edge degrees are
distributed according to a power law). This explains why few nodes
(larger circles) have many connections (darker lines), while many nodes
(small dots) have fewer connections (lighter lines). Note that the
minimum connectivity in our simulations is 3 (i.e., each node has at
least three neighbors and the maximum connectivity scales with the square
root of the number of nodes). Data for $N = 512$ nodes.
}
\label{fig:network}
\end{figure}

Networks play an integral role in all fields of science, as well as
numerous industrial applications. Virtually any process or group of
interacting entities can be described by a network.  In particular,
the study of {\em scale-free} networks (i.e., networks where the
degree distribution follows a power law) found a renewed interest
in the last decades after it had been shown that the internet follows
such a network topology \cite{albert:99}. Since then, there have been
numerous studies showing that a multitude of networks ranging from
computer networks, to protein-protein interaction networks, semantic
networks and, in particular social networks such as citation networks
or sexual partner networks are also well-described by scale-free
networks \cite{albert:02}.

Scale-free networks have edge degrees $\{k_i\}$ distributed according to
a power law $\lambda$, with the probability $\wp_k$ for a node to have
$k$ neighbors satisfying
\begin{eqnarray}
\label{eq:p_k}
\wp_k &\propto& k^{-\lambda}.
\end{eqnarray}
A typical network is shown in Fig.~\ref{fig:network}: While few nodes
have many edges connecting them to other nodes, many nodes have few
edges; the distribution of these following a power law.  Although
there have been several studies of Boolean variables on scale-free
networks with social interaction networks in mind, most studies
have focused on ``friend'' networks such as, for example, the Facebook
\cite{comment:facebook} network where person $A$ can ``friend'' person
$B$. Friendship can then be defined via a network edge between $A$
and $B$ with a {\em positive weight}.  However, other networks exist
where two persons $A$ and $B$ can either be ``friends'' or ``foes'' (i.e.,
a network with both positive- and negative-weight bonds). This appears,
for example, in the slashdot network \cite{comment:slashdot} or when
studying the robustness of opinion formation in, for example, an election
process. The latter type of network is rarely studied, possibly due
to the difficulties introduced by the negative-weighted edges in
the system.  However, they find wide applicability to many fields of
science such as the aforementioned social networks, as well as other
applications such as interaction networks between proteins or genes.

Why are Boolean problems on these scale-free networks interesting?
Because they can be seen as the simplest model to study how general
consensus forms on such a network for a decision problem with two
possible outcomes. By placing Boolean variables on each node of the
system, one can study either equilibrium or nonequilibrium properties
of the thermodynamics of the Boolean variables and thus see how stable
a given state of the system is.  Generalizations to more complex
decision problems can be readily accomplished by replacing the Boolean
variables with, for example, $q$-state Potts variables \cite{yeomans:92}.

The entities interacting on a real network may have complex
interactions, but models typically focus on connectivity in
randomly occupied networks or on networks with uniform interactions
between the entities. As illustrated above and as suggested in
different studies \cite{antall:06,leskovec:10}, many real networks
possess both friend and foe interactions among the degrees of freedom.
Because the network intrinsically has loops, this leads to frustration
between the Boolean variables, quickly complicating the study of
such systems.

Here, we study the critical behavior of the random-bond Ising model on
scale-free networks. The model maps directly onto a friend or foe network
(random bonds) with Boolean variables (Ising spins).  Crucially, both
ferromagnetic and antiferromagnetic interactions are allowed.  Although
many of the networks of experimental importance are dynamic and
out-of-equilibrium, a thorough understanding of the equilibrium model
provides a first step into the understanding of generic problems
associated with networks with random interactions.

Monte Carlo simulations of Ising spins on scale-free networks and
complex random graphs with {\em uniform antiferromagnetic} interactions
\cite{bartolozzi:06,herrero:09,weigel:07} have shown that a stable
spin-glass phase exists. Similarly, studies of a random-field {\em
ferromagnetic} Ising model \cite{imry:75,sethna:93} on scale-free
graphs \cite{lee:06a} show that for $\lambda \le 3$  the spins are
always ordered (i.e., consensus is stable to local perturbations),
whereas for $\lambda > 3$ a phase transition between a paramagnetic and
ferromagnetic phase exists as a function of the random-field strength.

Surprisingly, for the case of a pure Ising spin glass defined on a
scale-free graph no detailed numerical results exist with most results
relying on analytical approximations and mean-field calculations
\cite{mooij:04,kim:05,ferreira:10,ostilli:11}.  The detailed mean-field
study by Kim {\em et al.}~\cite{kim:05} showed that for $\lambda \le
3$ the critical temperature of the system diverges (i.e., the spins
are stable to arbitrary local perturbations), whereas for $\lambda >
3$ a finite-temperature transition from a paramagnetic to a spin-glass
state exists.  In this work we {\em improve} on the results by Kim {\em
et al.}~by expanding the approach of Leone {\em et al.}~\cite{leone:02}
for ferromagnetic systems to spin glasses. We present analytical
results backed up by numerical results (using large-scale Monte Carlo
simulations) for both Gaussian-distributed and bimodal edge weights
between the Ising spins. We show that when the probability $p$ of
positive (ferromagnetic) interactions and the probability $1-p$ for
negative (antiferromagnetic) interactions is the same, the system
undergoes a finite-temperature spin-glass transition if $\lambda >
3$, in agreement with previous results \cite{kim:05}. However, when
$\lambda \le 3$, a spin-glass (SG) phase is stable for all temperatures.
Finally, in the cases where both spin-glass and ferromagnetic (FM) order
would be expected ($p >
0.5$), only spin-glass order is present.  Relating back to the social
Gedankenexperiment, this would suggest that for certain networks the
opinion of the individual is robust towards local perturbations and
cannot be affected by global consensus.

In addition, we show that spin glasses on scale-free networks seem
to obey universality. This means that, in the pure spin-glass case,
the type of the interaction does not seem to affect the nature of the
order when opinion forms. Furthermore, for $\lambda > 4$ \cite{kim:05}
spin glasses on scale-free networks have the same universality class
as the mean-field Sherrington-Kirkpatrick \cite{sherrington:75}
Ising spin glass \cite{comment:fm}.

In Sec.~\ref{sec:model} we introduce the Hamiltonian studied, followed
by how the networks are constructed in Sec.~\ref{sec:graphs}. In
Sec.~\ref{sec:analytical} we present analytical results and construct a
$\lambda$--$p$ (network strength versus fraction of ferromagnetic bonds)
phase diagram. Details about the simulations are shown in
Sec.~\ref{sec:mc}, followed by numerical results in Sec.~\ref{sec:mcres}
and concluding remarks.

\section{Model}
\label{sec:model}

The Hamiltonian of the Edwards-Anderson Ising spin
glass \cite{edwards:75,binder:86} defined on a scale-free graph is
given by
\begin{equation}
{\mathcal H}(\{s_i\})=-\sum_{i<j}^N J_{ij} \varepsilon_{ij} s_i \,s_j , 
\label{eq:ham}
\end{equation}
where the Ising spins $s_i \in\{\pm 1\}$ lie on a scale-free graph
with $N$ sites and interactions 
\begin{eqnarray}
\label{eq:bond_disorder}
{\mathcal P}(J_{ij}, \varepsilon_{ij})
= \wp_J(J_{ij}) 
\left[  \Big(1-\frac{\kappa}{N} \Big) \delta( \varepsilon_{ij})
+ 
\frac{\kappa}{N} \delta( \varepsilon_{ij}-1) \right]. 
\end{eqnarray}
In Eq.~(\ref{eq:bond_disorder}) $\varepsilon_{ij}=1$ if a bond
is present between spin $s_i$ and $s_j$ and $\varepsilon_{ij}=0$
otherwise. $\kappa$ denotes the mean connectivity of the underlying
graph. The connectivity of site $i$, $k_i:=\sum_j \varepsilon_{ij}$,
is sampled from the scale-free distribution, Eq.~(\ref{eq:p_k}).
The bond values are drawn from either a Gaussian distribution with
zero mean and standard deviation unity, that is, 
\begin{equation}
\wp_J(J_{ij}) \sim \exp{(-J_{ij}^2/2)} 
\label{eq:gauss}
\end{equation}
or a bimodal distribution defined via
\begin{equation}
\wp_J(J_{ij}) =  (1-p)\delta(J_{ij} + 1) + p\delta(J_{ij} - 1) \, .
\label{eq:bim}
\end{equation}
In Sec.~\ref{sec:graphs} we describe in detail how the scale-free
graphs as shown in Fig.~\ref{fig:network} are generated for the
simulations.

\section{Generating scale-free graphs}
\label{sec:graphs}

One standard approach for the generation of scale-free networks is
preferential attachment \cite{barabasi:99}. In this physically inspired
growth process, a new node is added to the graph at each step, and the
probability of attaching to previously existing nodes depends on their
edge degrees.  This is believed to mimic the creation of scale-free
networks in a wide variety of processes, where newcomers are more
likely to associate with already-popular members of the network.
The frustration that must be present for the disordered problem to
be nontrivial requires that there be loops present in the network.
The new nodes must therefore attach to multiple pre-existing nodes
in this particular growth process.  The simplest implementation
of preferential attachment, where the probability of attaching
to a node of edge degree $k$ is proportional to $k$ produces a
power-law distribution of edge degrees with exponent $\lambda=3$.
It is possible to modify the exponent, at least in the $N\to \infty$
limit, by changing the function giving the probability of attachment
\cite{krapivsky:00, krapivsky:01}.

Another technique for generating scale-free networks is to extend
the classical ``configuration model'' \cite{bender:78, newman:03a,
catanzaro:05}.  Here, an edge degree distribution is chosen according
to a power law of exponent $\lambda$ and a graph is chosen randomly
from the ensemble of all possible graphs consistent with the chosen
edge degree distribution.  The chosen graph is then fixed in time for
a given sample (i.e.,~it is a quenched random graph).  The procedure
for generating the graphs starts by assigning $k$ stubs for each node,
where $k$ is drawn from the distribution $\wp_k$ [Eq.~(\ref{eq:p_k})],
and randomly pairing the stubs.  These pairings make up the edges
in the graph.  If the resulting graph is valid (in our case, we do
not allow double edges, and only connected graphs are considered),
then it is accepted and may be used for simulation.

In practice, we use a slightly different approach which is much
faster but is known to cause the selection of graphs to be slightly
nonuniform \cite{klein-hennig:11,comment:akh}.  If, during the pairing
process, a connection is to be made between two stubs corresponding to
the same node, this is not allowed.  In the method described above,
all edges are removed and the pairing starts from the beginning.
Here, we simply reject the pairing and move on.  This is not expected
to affect our results significantly: The degree distribution is
fixed independently of this method.  In practice, the preferential
attachment graphs are quite different than these random 
graphs, yet our tests give qualitatively similar results for the two
cases.  The results presented in this paper are from simulations
using the quenched random graphs as defined above.

The graph-generation technique used in our simulations works for
general degree distributions, although the acceptance rate may be
prohibitively low for some graphs.  For application to scale-free
graphs, an upper bound is imposed on the allowed edge degrees,
$k_\mathrm{max} = \sqrt{N}$.  Although we can generate graphs with $k$
exceeding $\sqrt{N}$, the ensemble is poorly defined in this case: Even
randomly chosen graphs cannot be uncorrelated
\cite{burda:03,boguna:04,catanzaro:05}. We also set a lower bound on the
edge degree $k_\mathrm{min}=3$. This eliminates spins which could be
easily integrated out of the system and do not contribute to the
frustration properties: dangling spins with only one attachment, and
(possibly long) loops of spins which are not connected with any other
spins.

\section{Analytical results}
\label{sec:analytical}

Analytical results for spin glasses on scale-free networks were
obtained previously \cite{kim:05}, and here we adapt a calculation
for the Ising ferromagnet on a scale-free graph to the spin-glass
case \cite{leone:02}.

\subsection{Replica approach}

We use the replica approach and at first consider the disorder average
of the replicated partition function $Z^n$ for integer powers of $n$
\begin{eqnarray} 
\label{eq:Z^n}
\lglo Z^n \rglo
&=&
\blglo \sum_{ \{  s^ a_i\}} 
\exp \left(- \beta \sum_{a=1}^n {\cal H}(\{s^a _i\}) \right) \brglo \\  \nonumber
&&
\hspace{-1cm}=
\int  \prod_{i<j}  \frac{ dJ_{ij}d\varepsilon_{ij} }{\mathcal{N}} 
P(J_{ij}, \varepsilon_{ij}) 
\prod_{j=1}^N \delta\Big( \sum_{j (\neq i) } \varepsilon_{ij} -k_j \Big ) 
Z^n  \, .
\end{eqnarray} 
The double angular brackets $\lglo \cdots \rglo$ denote an average only
over the quenched interaction variables  $J_{ij}$ and
$\varepsilon_{ij}$. At the moment the connectivities $k_i$ are fixed.
The constraints on these quantities, which are necessary to impose a
scale-free degree distribution, will be introduced later [see
Eq.~(\ref{eq:k-average})] as done byt the authors of Ref.~\cite{leone:02}.  In
Eq.~(\ref{eq:Z^n}) $\mathcal{N}$ is a normalization constant and $\beta
= 1/T$ the inverse temperature.  After an appropriate continuation to
small values of $n$ the expression in Eq.~(\ref{eq:Z^n}) is related to
the free energy  $F$ of the model via
\begin{eqnarray} 
\beta F= -\lim_{n \to 0} \frac{\lglo { Z^n}\rglo -1}{n} \, .
\end{eqnarray} 
Using a representation of the $\delta$-function, the integrals in
Eq.~(\ref{eq:Z^n}) factorize resulting in
\begin{eqnarray} 
\label{eq:Z^n_2}
\lglo Z^n \rglo \!\!\! &=& \!\!\!
\sum_{ \{ \vec s_i\}}\frac{e^{-\kappa N/2}}{ \mathcal{N}}
\int \prod_i \left(\frac{d {\psi}_i}{2 \pi}\right)
\exp \left(-i \sum_{i=1}^N \psi_i k_i\right) 
\\ \nonumber && \hspace{+1.3cm} \times 
\exp \left[ \frac{\kappa}{2N} \sum_{ij}  \left \langle 
e^{\beta J \vec s_{i} \cdot \vec s_{j}}
\right \rangle_J e^{  i \psi_{i}} e^{i \psi_{j}}\right] ,
\end{eqnarray} 
where $\vec s_i$ is an Ising spin with $n$ components.  The order
parameters
\begin{equation}
\label{eq:OP_Leone}
 \rho (\vec{\sigma}) =
\frac{1}{N} \sum_i \delta (\vec{\sigma},{\vec{s}}_i) e^{i \psi_i} ,
\end{equation} 
and their conjugated fields $\hat \rho(\vec \sigma)$ allow one to
perform the trace over the spin variables $ \vec s_i $ in
Eq.~(\ref{eq:Z^n_2}).  After replacing the connectivity-dependent site
averages with the appropriate averages over the degree distribution
$\wp_k$ we obtain
\begin{eqnarray} 
\nonumber
\label{eq:k-average}
\frac{1}{N} \sum_{i=1}^N 
\log \sum_{\vec \sigma} \left( \hat \rho( \vec \sigma)\right)^{k_i}
&\equiv& 
\sum_{k} \wp_k \log \sum_{\vec \sigma} 
\left( \hat \rho( \vec \sigma) \right)^{k} 
\\ 
&=&\beta f_{\lambda}(\hat \rho) .
\end{eqnarray}
The partition function $\lglo Z^n \rglo$ then acquires the form 
\begin{eqnarray} 
\label{eq:Z^n_saddle_point}
\lglo Z^n \rglo 
\propto  
\int \prod_{\vec \sigma} d \rho(\vec{\sigma})  d \hat{\rho}(\vec{\sigma})
\exp \Big(- N \beta f_{\mathrm{trial}} (  \rho , \hat \rho)  \Big)
\end{eqnarray}
with the trial free energy
\begin{eqnarray} 
\label{eq:f_trial}
\beta f_{\mathrm{trial}}(\rho ,\hat \rho)
&=& 
\kappa \sum_{\vec{\sigma}}\rho (\vec{\sigma}) \hat{\rho}(\vec{\sigma}) 
+ 
\frac{\kappa}{2} - \beta f_{\lambda}(\hat \rho) \\ \nonumber
&& \hspace{-0.2cm} - \frac{\kappa}{2} \sum_{\vec\sigma,\vec\tau} 
\rho (\vec\sigma)\rho (\vec\tau)  \left \langle 
e^{\beta J (  \vec \sigma,  \vec \tau )}\right \rangle_J ,
\end{eqnarray} 
where the average over the distribution $\wp_J$ is  represented as
$\langle \cdots \rangle_J$.  The integral in
Eq.~(\ref{eq:Z^n_saddle_point}) can be evaluated with the method of
steepest descent leading to a self-consistent equation for the order
parameters $\rho$:
\begin{eqnarray}
\label{eq:SP}
{\rho}(\vec{\sigma}) =  
\sum_k \frac{k}{\kappa}\, \wp_k  \,
\left(
\sum_{\vec\tau}  \rho (\vec\tau) 
\left \langle e^{\beta J (  \vec \sigma,  \vec \tau ) }\right \rangle_J
\right )^{k-1} .
\end{eqnarray} 
Finding stable solutions to this equation in the \mbox{$n
\to 0 $} limit would lead to the free energy of the model at
all temperatures. This problem  remains to be solved for general
spin-glass Hamiltonians. However, here we are interested {\em only} in
the transition temperature of the system.  In this case the simplest
replica symmetric analysis is sufficient.

\subsection{Replica-symmetric solution}

Due to the Boolean nature of the Ising spins, the replica-symmetric
solution ${\rho}_{ \mathrm{rs}}(\vec \sigma)$ of Eq.~(\ref{eq:SP})
only depends on the sum $s=\sum _a \sigma_a$ of the components of
$\vec \sigma$.  The parametrization
\begin{eqnarray}
\label{eq:RS}
 {\rho}_{ \mathrm{rs}}(\vec{\sigma})= \int dh \mathcal{P}(h) \exp(\beta h s)  , 
\end{eqnarray} 
allows one to perform the limit \mbox{$n \to 0$} straightforwardly,
which leads to the self-consistent equation for the local-field
distribution
\begin{eqnarray}\nonumber
\label{eq:P(h)}
\mathcal{P}(h)&=&
\sum_{k=k_{ \mathrm{min}}}^\infty 
\wp_k \frac{k}{ \kappa}  \int \prod_{i=1}^{k-1} d h_i d J_i 
\mathcal{P}(h_i)\wp_J(J_i)  \\
&& \hspace{+2.1cm}
\times \delta \left( h- \sum_{i=1}^{k-1} u(h_i, J_i) \right) 
\end{eqnarray}
with $u$ defined as 
\begin{equation}
u(h, J)=\frac{1}{\beta}
\mathrm{atanh}[\tanh(\beta h)\tanh( \beta J)] .
\end{equation}

As pointed out in Ref.~\cite{leone:02} and calculated in
Ref.~\cite{wemmenhove:05} this equation can be derived within the
cavity framework \cite{mezard:01}. When connecting a new site to the
system, one has to take into account the heterogeneity of the graph,
which is reflected in the distribution $(k/\kappa)\wp_k$ on the
right-hand side of Eq.~(\ref{eq:P(h)}).  The spin-glass order parameter
\mbox{$q=\lglo s_i\rangle_T^2\rangle_J$}, where $\langle \cdots \rangle_T$
represents a thermal average, is related to the local-field distribution
via
\begin{eqnarray}
q&=&\sum_{k=k_{ \mathrm{min}}}^\infty \wp_k  \int \prod_{i=1}^{k} 
d h_i d J_i \mathcal{P}(h_i)\wp_J(J_i) \\ \nonumber
&& \hspace{+1.8cm} \times
\tanh^2 \left( \beta \sum_{i=1}^k u(h_i,J_i) \right), 
\end{eqnarray}
whereas the magnetization $m=\lglo  s_i\rangle_T \rangle_J$ is
given by
\begin{eqnarray}
m&=&\sum_{k=k_{ \mathrm{min}}}^\infty \wp_k  
\int \prod_{i=1}^{k} d h_i d J_i \mathcal{P}(h_i)\wp_J(J_i) \\ \nonumber
&& \hspace{+1.8cm}\times
\tanh \left( \beta \sum_{i=1}^k u(h_i,J_i) \right) .
\end{eqnarray} 
The expressions for the order parameters are derived by using real
replicas of the system.  Following the notation of Viana and Bray we
also introduce the quantities
\begin{eqnarray}
\label{eq:q_n}
q_n=\int dh \, \mathcal{P}(h)  \tanh^n \left( \beta h \right),
\end{eqnarray}
and remind the reader, that the inequality  $q_2 \geq q_n$ holds for all
even $n>2$. 

The $\delta$-function is always a solution of the self-consistent
equation, Eq.~(\ref{eq:P(h)}). Due to the vanishing of all local
fields, and consequently of the order parameters $m$ and $q$, this
solution intuitively corresponds to the paramagnetic phase. Based
on physical grounds, we expect this solution to be unstable below
a critical temperature, which thus signals a transition to a frozen
low-temperature phase.

We first test the stability of this paramagnetic solution towards
a spin-glass transition. To this end we introduce a distribution
$\mathcal{P}_0$ with a small $q_2=\epsilon$ on the right-hand
side of Eq.~(\ref{eq:P(h)}). If the corresponding quantity of the
resulting distribution on the left-hand side exceeds $\epsilon$,
the paramagnetic solution becomes unstable and the system undergoes
a spin-glass transition. The critical temperature $T^{\mathrm{SG}}_c$
is given by the equation
\begin{eqnarray}
\label{eq:Tc_SG}
 1 = \sum_k \frac{k(k-1)}{ \kappa} \wp_k \left \langle\tanh^2
 \left(\frac{J}{ T^\mathrm{SG}_c	
}\right )\right \rangle_J\, .
\end{eqnarray}
To detect a transition towards a ferromagnetic phase a similar
procedure using a distribution $\mathcal{P}_0$  with a small
$q_1=\epsilon$ can be applied. This leads to the stability criterion
to determine $T^\mathrm{FM}_c$, that is,
\begin{eqnarray}
\label{eq:Tc_FM}
1 = \sum_k \frac{k(k-1)}{ \kappa} \wp_k \left \langle \tanh
\left(\frac{J}{ T^\mathrm{FM}_c}\right )\right \rangle_J\, . 
\end{eqnarray}

When lowering the temperature the system enters the frozen phase with
the higher $T_c$. Within the low-temperature phase no conjectures can
be made from this ``paramagnetic'' analysis.  Note that the usual
procedure to determine $T_c$ (which relies on the moments of the
distribution  $\mathcal{P}$) fails here for small values of the
scale-free decay parameter ($\lambda \leq 4$), because in this region
the moments cease to exist.

\begin{figure}[tb] 
\includegraphics[width=\columnwidth]{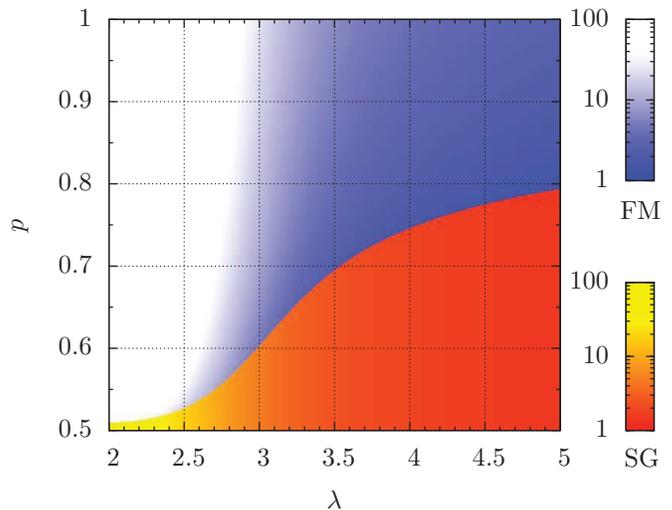}
\caption{(Color online) 
Analytical phase diagram: Fraction of ferromagnetic bonds $p$
vs exponent $\lambda$.  Color represents the value of the critical
temperature. The darker the color, the smaller the numerical value.  
For $\lambda
\le 3$, both $T_c^{\mathrm{FM}}$ and $T_c^{\mathrm{SG}}$ diverge,
but there is a crossover in the rates at which they diverge. In this
case, in the blue-white region (top shaded region of the graph)
the system is ferromagnetic at all finite temperatures, while in
the red-yellow region (bottom shaded region of the graph), the
system is a spin glass at all finite temperatures.  Note that for
$\lambda \to 2$ the system becomes increasingly stable against local
(temperature) perturbations.  For the particular case of $p = 0.5$
(horizontal axis) $T_c^{\mathrm{SG}} \to \infty$ for $\lambda \le 3$
and $T_c^{\mathrm{SG}}$ finite for $\lambda > 3$.
} 
\label{fig:phase_diagram}
\end{figure}

For a bimodal bond distribution [Eq.~(\ref{eq:bim})] with
minimum connectivity \mbox{$k_{ \mathrm{min}}=3$} the solutions
of Eqs.~(\ref{eq:Tc_SG}) and (\ref{eq:Tc_FM}) are visualized
in Fig.~\ref{fig:phase_diagram}. The color represents the value of
the critical temperature. The darker the color, the smaller the numerical
value. Note that both transition temperatures diverge for $\lambda
\leq 3$, because the second moment of the degree distribution is
infinite. In the blue-white region (top shaded region of the graph)
the system is always ferromagnetic with $T_c^{\mathrm{FM}}$ increasing
for $\lambda \to 2$. In the red-yellow region (bottom shaded region of
the graph) the system is a spin glass at all finite temperatures with
$T_c^{\mathrm{SG}}$ increasing for $\lambda \to 2$. For the particular
case of $p = 0.5$ $T_c^{\mathrm{SG}} \to \infty$ for $\lambda \le 3$
and $T_c^{\mathrm{SG}}$ finite for $\lambda > 3$.

\subsection{Comparison to the static model}

We briefly compare our result to previous calculations.  In the work of
Kim {\em et al.}~\cite{kim:05} the ``static'' model was used.  In this
approach a probability $\pi_i \propto {i^{-\mu}}$, $\mu \in [0,1)$ is
assigned to each vertex $i = 1,  \ldots,  N$ to obtain a scale-free graph
with degree $\lambda=1+\mu^{-1}$ and mean connectivity $\kappa$. The
results derived for the phase boundaries in Ref.~\cite{kim:05}
rely on a {\em truncation scheme} which confines the number of the
order parameters to the ``most important'' ones \cite{viana:85}. At
the replica-symmetric level, this approximation is equivalent to the
assumption of Gaussian local fields, which can be avoided, using a
variant of the approach described above.

Starting from Eq.~(10) in Ref.~\cite{kim:05} we proceed using
the order parameters
\begin{eqnarray}
\rho (\vec{\sigma}) =
\frac{1}{N} \sum_i  \pi_i \delta (\vec{\sigma},{\vec{s}}_i) \,
\end {eqnarray}
and their conjugated variables $\hat \rho(\vec \sigma )$.  The
functions which still depend on the vertex weights $\pi_i$ can be
transformed to integrals in the thermodynamic limit according to
\begin{eqnarray}
\nonumber
\frac{1}{N}\sum_{i=1}^N g( \kappa N \pi_i) \approx \int_0^1
\hspace{-0.2cm} dx g \left( (1-\mu) \frac{\kappa}{x^\mu} \right)= 
\int_{k_{ \mathrm{min} }}^\infty \hspace{-0.4cm} d k \, \wp(k) g(k) \, .
\end{eqnarray}
In the last step we performed a substitution which directly leads
to the scale-free distribution relevant for the static model
\begin{eqnarray}
\label{eq:p(k)_Kim}
\wp(k)&=& 
\big(k_\mathrm{min}\big)^{\lambda-1} \frac{\lambda-1}{k^\lambda} , 
\quad
k_\mathrm{min}=\frac{\lambda-2}{\lambda-1} \kappa \,.
\end{eqnarray}
The partition function $\lglo Z^n \rglo$ of
the static model reduces to a saddle-point integral
[see Eq.~(\ref{eq:Z^n_saddle_point})] with a scale-free dependent
part in the the trial free energy (\ref{eq:f_trial}):
\begin{eqnarray} 
\nonumber
\beta f_{\lambda}( \hat \rho) 
&=& 
\int_{k_{ \mathrm{min} }}^\infty \hspace{-0.4cm} d k \, \wp(k) 
\log \left[\sum_{\vec{\sigma}} 
\exp\left( k \hat\rho(\vec{\sigma})\right)\right]\, .
\end{eqnarray} 
Using the saddle-point equations and the replica-symmetric Ansatz, we
obtain a self-consistent equation for the distribution of local fields
\begin{eqnarray}
\label{eq:P(h)_Kim}
\mathcal{P}(h) 
& = & \int_{k_\mathrm{min}}^\infty \hspace{-0.4cm} d k \, \wp(k)  
\frac{k}{\kappa} \sum_{m=0}^{\infty}\frac{e^{-k}k^m}{m!}  
\int  \prod_{i=1}^{m} dh_i d J_i \\ \nonumber
&&\hspace{+1cm} \times\mathcal{P}(h_i)\wp_J(J_i)  
\delta \left(h- \sum_i^{m} u(h_i,J_i)\right)\,.
\end{eqnarray} 
This equation is a generalization of the replica-symmetric equations
for the Viana-Bray model \cite{viana:85,monasson:98} where the mean
connectivity $k$ is sampled from the distribution $(k/\kappa)\wp(k)$.
Moreover, the equation is similar to the self-consistent equation
[Eq.~(\ref{eq:P(h)})], where the connectivities are sampled
from the distribution $(k/\kappa)\wp_k$.  This last approach is
a generalization of the fixed-connectivity model to arbitrary
degree distributions, and we prefer it due to its generality.
The paramagnetic solution $\mathcal{P}(\cdot)=\delta(\cdot)$ is a
solution of Eq.~(\ref{eq:P(h)_Kim}) which becomes unstable towards
a spin-glass transition at the critical temperature $T^\mathrm{SG}_c$
\begin{eqnarray}
\nonumber
\label{eq:Tc_SG_Kim}
1 &= & \left \langle\tanh^2 \left(\frac{J}{T^\mathrm{SG}_c}\right)
\right \rangle_J \int_{k_\mathrm{min}}^\infty d k \, \wp(k) 
\frac{k^2}{\kappa}\, \\
  &=& \kappa \left \langle\tanh^2 \left(\frac{J}{ T^\mathrm{SG}_c}\right )  
\right \rangle_J 
 \frac{(\lambda-2)^2}{(\lambda-1)(\lambda-3)} \,.
\end{eqnarray}
The transition to a ferromagnetic phase  occurs at the critical
temperature $T^\mathrm{FM}_c$
\begin{eqnarray}
 1 &= &\kappa \left \langle
 \tanh \left(\frac{J}{ T^\mathrm{FM}_ c	}\right ) 
 \right \rangle_J
   \frac{(\lambda-2)^2}{(\lambda-1)(\lambda-3)} \,.
\end{eqnarray}
In the last line of Eq.~(\ref{eq:Tc_SG_Kim}) we performed
the integrals which are convergent for $\lambda> 3$ {\em only}.
This representation coincides with the predictions  for the critical
temperature in Ref.~\cite{kim:05}. Here the results were
obtained {\em without resorting to a truncation scheme}. Predictions
for the de Almeida Thouless line \cite{almeida:78} relying on Gaussian
approximations \cite{cizeau:93} can lead to results which were shown to
be wrong \cite{neri:10,janzen:10a}. The investigation of the phase
boundaries in the presence of an external magnetic field seems
promising in this approach, and will be reported elsewhere.

\subsection{Critical exponents}

We now turn to the computation of the critical exponent which
governs the growth of the order parameter close to the transition
temperature. The order parameter is proportional to $\tau^{\beta}$,
here $\tau$ is the reduced temperature (i.e., $\tau= 1-{T}/{T_c}$
and $\beta$ a critical exponent).

Note that close to $T_c$ all $q_n$ are small, as $\mathcal{P}$ is
close to a $\delta$ function. This allows us to neglect $q_n$ with
large $n$ in this region while keeping the dominant terms (i.e., 
$q_2$ for the spin-glass transition and $q_1$ for the ferromagnetic
transition). In particular, close to $T_c$ the order parameters $q$
and $m$ are proportional to $q_2$ and $q_1$, respectively. It is
therefore sufficient to investigate how these quantities evolve from
zero below $T_c$.

We start with the spin-glass transition and derive a self-consistent
equation for $q_2$:
\begin{eqnarray}
\label{eq:q2}
q_2&\approx & \sum_{k} \wp_k\frac{k}{\kappa}  \sum_{l=1}^{k-1} t_{l} 
 {k-1 \choose l}  q_2^l \langle  \tanh^{2}(\beta J)\rangle^l_J,
\end{eqnarray}
where $t_l$ are Taylor coefficients of $\tanh^2$. To
obtain Eq.~(\ref{eq:q2}) we use the self-consistent equation
[Eq.~(\ref{eq:P(h)})] for $\mathcal{P}$ and employ a series expansion
in terms of the $\{\tanh(\beta h_i)\tanh(\beta J_i)\}$. We then
evaluate all averages with respect to $\mathcal{P}$ and $\wp_{J}$
and finally perform the aforementioned approximation (i.e., neglect
all $q_n$ with $n>2$).

If $\lambda > 4$ we recover the Sherrington-Kirkpatrick mean-field
critical exponents by truncating the second sum after the $l=2$
contribution on the right hand side of the equation, which amounts to
\begin{eqnarray}
\label{eq:betaSK}
 \nonumber
q \propto \frac{1 - \langle k(k-1) \rangle_k \kappa^{-1}  
\left \langle \tanh^2 \left(  \beta J \right ) 
\right \rangle_J }{ \langle k(k-1)(k-2) 
\rangle_k \left \langle \tanh^2 \left(  \beta J \right ) 
\right \rangle_J^2 } \propto  \tau+ \mathcal{O}(\tau^2),
\end{eqnarray}
that is, 
\begin{equation}
\beta = 1 , 
\;\;\;\;\;\;\;\;\;\;\
\;\;\;\;\;\;\;\;\;\;\
(\lambda > 4).
\end{equation} 
For $\lambda \leq 4$ the $k$ average in the denominator of the last
equation diverges and the usual technique to extract the critical
exponent does not work.

For $\lambda \leq 4$ we note that due to the combinatorial factor
in Eq.~(\ref{eq:q2}) the lower (important) powers $l$ of $q$ have
a prefactor proportional to $k^l$, such that the divergent part
\mbox{($l \geq 2$)} depends on the combination $k q_2 \left \langle
\tanh^2(\beta J) \right \rangle_J$ only. We assume that this is the
important $k$ dependence and introduce a function $F$ in the following
way:
\begin{eqnarray}
\nonumber
&& \sum_{k} \wp_k\frac{k}{\kappa} 
   \sum_{l=2}^{k-1} t_l {k-1 \choose l}    
    q_2^l \left \langle  \tanh^2(\beta J) \right  \rangle^l_J \\ \nonumber
&&  \hspace{+0.5cm} \approx \sum_{k} \wp_k\frac{k}{\kappa} 
    F\left(kq_2 \left \langle  \tanh^2(\beta J) \right  \rangle_J\right)  
    \\ \nonumber
&&  \hspace{+0.5cm}\approx  \int_{k_{\rm min}}^{\infty} dk  
    \frac{c}{ \kappa}\frac{1}{ k^{\lambda-1}} 
    F\left(kq_2\left \langle  \tanh^2(\beta J) \right  \rangle_J \right) 
    \\ \nonumber
&& \hspace{+0.5cm}\approx
    \left(q_2\left \langle  
        \tanh^2(\beta J) 
    \right  \rangle_J \right)^{\lambda-2}
    \int_{x_{\rm min} }^{\infty}  dx  
    \frac{c}{ \kappa}\frac{1}{ x^{\lambda-1}} F(x).
\end{eqnarray}  
In the last line we scaled out $q_2$ by a substitution leading to the
lower integration bound \mbox{$x_{\rm min}= q_2\left \langle \tanh^2(\beta
J) \right \rangle_J k_{\rm min}$}.  The quadratic dependence of $F$
at the origin due to the fact that the series expansion starts with $l = 2$ 
allows one to put $x_{\rm min} \propto q_2 \to 0$ for $\lambda
<4$. Inserting the above steps into Eq. (\ref{eq:q2}) we obtain
\begin{eqnarray}
\nonumber
\label{eq:beta_non_SK}
\nonumber q^{\lambda-3} \propto \frac{ 
1 - \langle k(k-1) \rangle_k \kappa^{-1}  
\left \langle \tanh^2 \left(  \beta J \right ) \right \rangle_J  }
{\left \langle  \tanh^2(\beta J) \right \rangle_J ^{\lambda-2} 
\int_{0}^{\infty}  dx  \frac{c}{ \kappa}\frac{1}{ x^{\lambda-1}} F(x) } 
\propto  \tau + \mathcal{O}(\tau^2).
\end{eqnarray}
This means 
\begin{equation}
\beta= 1/(\lambda-3) ,
\;\;\;\;\;\;\;\;\;\;\
\;\;\;\;
(3 < \lambda < 4) ,
\end{equation} 
which agrees with the result of Kim {\em et al.}~\cite{kim:05} derived
within the static approximation.  The limiting case $\lambda=4$ needs
some special care, leading to logarithmic corrections [i.e., $q \propto
-\tau/\log(\tau)$].

To reproduce the results  for the critical exponent in the
ferromagnetic sector, which were calculated by the authors of 
Ref.~\cite{leone:02},
the same technique can be used. In particular
\begin{equation}
\beta = 1/2 ,
\;\;\;\;\;\;\;\;\;\;\
\;\;\;\;\;\;\;\;\;\;
(\lambda  > 5).
\end{equation}
This means the system is in the mean-field universality class for
$\lambda>5$ because, within $\mathcal{O}(\tau^2)$
\begin{eqnarray}
\label{eq:betaFM}
 \nonumber
m^2 \propto  
\frac{1- \langle k(k-1) \rangle_k \kappa^{-1}  
\left \langle \tanh \left(  \beta J \right ) 
\right \rangle_J  }{ 
\langle k(k-1)(k-2)(k-3) \rangle_k \left \langle \tanh 
\left(  \beta J \right ) \right \rangle_J^3 } 
\propto  \tau+ \ldots .
\end{eqnarray}
For $\lambda \leq 5$ we face the same problem as in the spin-glass
sector when $\lambda \leq 4$, because the $k$ average in the denominator
diverges. Performing analogous considerations leads to
\begin{equation}
\beta= 1/(\lambda-3) , 
\;\;\;\;\;\;\;\;\;\;\
\;\;\;\;\;\;\;\;\;\;
(3 < \lambda  < 5).
\end{equation}
Finally, for $\lambda = 5$, $m^2 \propto -\tau/\log(\tau)$.

Summarizing, for $\lambda > 4$ [$\lambda > 5$] in the SG [FM] sector, the
critical exponents agree with the mean-field case, whereas for $3 <
\lambda < 4$ [$3 < \lambda < 5$] in the SG [FM] sector the critical
exponents depend on the exponent $\lambda$.

\section{Numerical details}
\label{sec:mc}

We validate the aforementioned analytical results using Monte Carlo
simulations. In particular, the numerical results show the strength
of the different corrections to scaling depending on the choice of
the exponent $\lambda$ of the scale-free network.

\subsection{Observables}

To determine the location of both the ferromagnetic and spin-glass
phase transitions we measure first the Binder cumulant \cite{binder:81}
defined via
\begin{equation}
g = \frac{1}{2}
\left(
3 - \frac{\lgle {\mathcal O}^4\rgle}{\lgle{\mathcal O }^2\rgle^2}
\right) \; .
\label{eq:binder}
\end{equation}
In Eq.~(\ref{eq:binder}), $\lgle \cdots \rgle$ represents an average
over the disorder via $\langle \cdots \rangle_J$, the random graphs via
$\langle \cdots \rangle_k$, and Monte Carlo time (i.e., $\langle \cdots
\rangle_T$). Furthermore, ${\mathcal O}$ is either the magnetization
$m$ in the ferromagnetic sector defined via
\begin{equation}
m = \frac{1}{N}\sum_i s_i
\label{eq:magn}
\end{equation}
or the spin-glass overlap $q$ in the spin-glass sector given by
\begin{equation}
q = \frac{1}{N}\sum_{i = 1}^N s_i^\alpha s_i^\beta \; .
\label{eq:q}
\end{equation}
In Eq.~(\ref{eq:q}), ``$\alpha$'' and ``$\beta$'' represent two
copies of the system with the same disorder.  The Binder ratio is
dimensionless and thus has the simple scaling form
\begin{equation}
g = \widetilde{G}\left(N^{1/\nu}[T - T_c] \right)  \; ,
\label{eq:g_scale}
\end{equation}
where $T_c$ represents the transition temperature. The expression in
Eq.~\eqref{eq:g_scale} is valid in the non-mean-field region (i.e.,
for $\lambda < 4$ in the spin-glass sector and $\lambda < 5$ in the
ferromagnetic sector \cite{kim:05}). In the spin-glass mean-field region
($\lambda > 4$) Eq.~\eqref{eq:g_scale} is replaced by \cite{larson:10}
\begin{equation}
g = \widetilde{G}\left(N^{1/3}[T - T_c] \right)  \; .
\label{eq:g_scale_mf}
\end{equation}
Note that the two-point finite-size correlation length
\cite{cooper:82,palassini:99b,ballesteros:00,martin:02} typically
used to pinpoint transitions in glassy systems cannot be used here
because the spins are placed on a lattice that has no geometry.

The calculation of the Binder ratio $g$ allows one to determine $T_{\rm
c}$ and the critical exponent $\nu$ for both the spin-glass and
ferromagnetic sectors. However, to fully characterize the critical
behavior of the model, a second critical exponent has to be computed
\cite{yeomans:92}. We have also computed the susceptibility $\chi$ given
by
\begin{equation}
\chi = N \lgle {\mathcal O}^2 \rgle \, .
\label{eq:chi}
\end{equation}
In the ferromagnetic case we therefore measure
\begin{equation}
\chi_m = N \lgle m^2 \rgle
\label{eq:chifm}
\end{equation}
with the magnetization $m$ given by Eq.~\eqref{eq:magn}, whereas in the
spin-glass case we measure
\begin{equation}
\chi_q = N \lgle q^2 \rgle
\label{eq:chisg}
\end{equation}
with the spin-glass overlap $q$ defined in Eq.~\eqref{eq:q}.
In general, the scaling behavior of the susceptibility is given by
\begin{equation}
\chi = N^{2 - \eta} \widetilde{C}
\left(N^{1/\nu}[T - T_c]\right) \, ,
\label{eq:chi_scale}
\end{equation}
where a simple finite-size scaling yields the exponent $\eta$.
Unfortunately, fluctuations in the data are huge and thus the
determination of the critical exponent $\eta$ is not possible.
However, in the mean-field regime, the finite-size scaling form
presented in Eq.~\eqref{eq:chi_scale} is replaced by
\begin{equation}
\chi = N^{1/3} \widetilde{C}
\left(N^{1/3}[T - T_c]\right) \, .
\label{eq:chi_scale_mf}
\end{equation}
Therefore, curves of $\chi/N^{1/3}$ should have the same scaling
behavior as the Binder ratio [Eq.\eqref{eq:g_scale_mf}]: When $T = T_c$
data for different system sizes cross at a point (up to scaling
corrections).

\subsection{Equilibration and simulation parameters}

\begin{figure}[tb]
\includegraphics[width=\columnwidth]{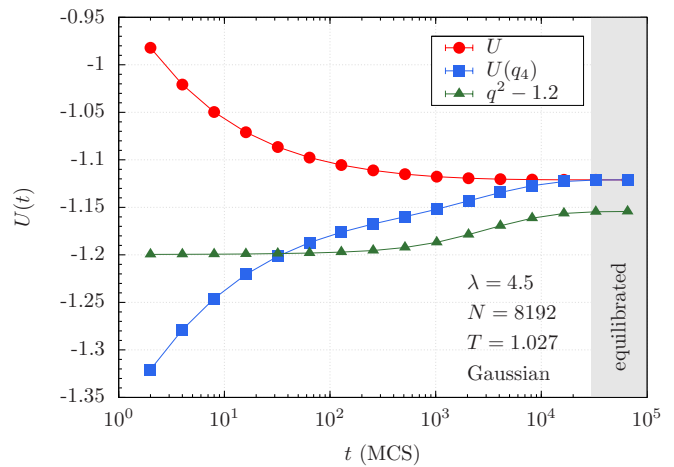}
\caption{(Color online)
Example equilibration plot for Gaussian disorder for $N = 8192$ spins at
$T = 1.027$ (lowest temperature simulated) and $\lambda = 4.5$.  Once
the data for the energy $U$ and the energy computed from $q_4$
[$U(q_4)$] agree, the system is in thermal equilibrium. This can be seen
also with data for $q^2$ that also are independent of Monte Carlo time.
Note that the data for $q^2$ are shifted by 1.2 for better viewing. Error 
bars are smaller than the symbols.
}
\label{fig:equil}
\end{figure}

The simulations are done using the parallel tempering Monte Carlo method
\cite{hukushima:96}. For the pure spin glass we first simulate the
system with Gaussian disorder to obtain an idea of the equilibration
behavior when the Ising model with disorder is defined on a scale-free
graph.  Furthermore, in the Gaussian case we can perform a rigorous
equilibration test \cite{katzgraber:01,katzgraber:09b} where the energy
per spin [$U = (1/N)\lgle {\mathcal H}\rgle$ with ${\mathcal H}$ defined
in Eq.~\eqref{eq:ham}] is compared to an expression derived from the
link overlap $q_4$ [defined below in Eq.~(\ref{eq:equil})].  The data
for both the energy per spin $U$ and the energy per spin computed from
the link overlap,
\begin{equation}
U(q_4) = - \frac{1}{T}\,\blgle \frac{N_b}{N}\, (1 - q_4)
                \brgle \, ,
\label{eq:equil}
\end{equation}
where
\begin{equation}
q_4 =  \frac{1}{N_b}  \sum_{i,j} \varepsilon_{ij}
        s_i^\alpha s_j^\alpha s_i^\beta s_j^\beta \, ,
\label{eq:q4}
\end{equation}
have to coincide when the system is in thermal equilibrium. In
Eqs.~\eqref{eq:equil} and \eqref{eq:q4} $N_b$ is the number of 
nonzero bonds of a given
sample.  Note that the expression in Eq.~(\ref{eq:equil}) is valid for
the spin-glass sector, however, it represents a conservative bound for
the ferromagnetic sector \cite{katzgraber:01}. Furthermore, $N_b$ is
{\em inside} the disorder average because the number of bonds fluctuates
from sample to sample.  Sample data are shown in Fig.~\ref{fig:equil}.
Once $U = U(q_4)$ the data for the squared order parameter $q^2$
(shifted for better viewing in Fig.~\ref{fig:equil}) are also in thermal
equilibrium.

For the bimodal disorder distribution and the ferromagnetic sector
the aforementioned equilibration test cannot be used. In this case
we preform a logarithmic binning of all observables. Once the data
for the last three bins agree within error bars we deem the system
to be in thermal equilibrium. The simulation parameters are shown in
Table \ref{tab:simparams}.

\begin{table}
\caption{
Parameters of the simulation: For each exponent $\lambda$, system size $N$ 
and fraction of ferromagnetic bonds $p$ (note: Gaussian disorder is marked 
with ``Gauss'') we compute $N_{\rm sa}$ disorder or network instances. 
$N_{\rm sw}
= 2^b$ is the total number of Monte Carlo sweeps for each of the $2 N_T$
replicas for a single instance, $T_{\rm min}$ [$T_{\rm max}$] is the
lowest [highest] temperature simulated, and $N_T$ is the number of
temperatures used in the parallel tempering method for each system size
$N$.
\label{tab:simparams}
}
\begin{tabular*}{\columnwidth}{@{\extracolsep{\fill}} c l l r l l l l }
\hline
\hline
$\lambda$ & $p$ & $N$ & $N_{\rm sa}$ & $b$ & $T_{\rm min}$ & $T_{\rm max}$ & $N_{T}$  \\
\hline
$3.0$  & $0.500$  & $1024$ & $10\,000$ & $15$ & $2.011$ & $4.208$ & $37$ \\
$3.0$  & $0.500$  & $2048$ & $ 9\,940$ & $15$ & $2.011$ & $4.208$ & $37$ \\
$3.0$  & $0.500$  & $4096$ & $12\,877$ & $17$ & $2.011$ & $4.208$ & $37$ \\
$3.0$  & $0.500$  & $8192$ & $ 5\,399$ & $19$ & $2.011$ & $4.208$ & $37$ \\
\hline
$3.5$  & $0.500$  & $1024$ & $20\,416$ & $15$ & $2.011$ & $4.208$ & $37$ \\
$3.5$  & $0.500$  & $2048$ & $10\,396$ & $15$ & $2.011$ & $4.208$ & $37$ \\
$3.5$  & $0.500$  & $4096$ & $18\,683$ & $17$ & $2.011$ & $4.208$ & $37$ \\
$3.5$  & $0.500$  & $8192$ & $12\,382$ & $19$ & $2.011$ & $4.208$ & $37$ \\
\hline
$4.5$  & $0.500$  & $1024$ & $ 9\,600$ & $16$ & $1.027$ & $2.410$ & $27$ \\
$4.5$  & $0.500$  & $2048$ & $ 9\,600$ & $16$ & $1.027$ & $2.410$ & $27$ \\
$4.5$  & $0.500$  & $4096$ & $ 9\,439$ & $19$ & $1.027$ & $2.410$ & $27$ \\
$4.5$  & $0.500$  & $8192$ & $ 9\,870$ & $19$ & $1.027$ & $2.410$ & $27$ \\
\hline
$4.5$  & $0.700$  & $1024$ & $ 9\,600$ & $16$ & $1.170$ & $3.949$ & $49$ \\
$4.5$  & $0.700$  & $2048$ & $ 8\,723$ & $16$ & $1.170$ & $3.949$ & $49$ \\
$4.5$  & $0.700$  & $4096$ & $10\,714$ & $16$ & $1.170$ & $3.949$ & $49$ \\
$4.5$  & $0.700$  & $8192$ & $ 8\,184$ & $17$ & $1.170$ & $3.949$ & $49$ \\
\hline
$4.5$  & $0.850$  & $1024$ & $13\,914$ & $16$ & $1.170$ & $3.949$ & $49$ \\
$4.5$  & $0.850$  & $2048$ & $12\,103$ & $16$ & $1.170$ & $3.949$ & $49$ \\
$4.5$  & $0.850$  & $4096$ & $ 9\,570$ & $16$ & $1.170$ & $3.949$ & $49$ \\
$4.5$  & $0.850$  & $8192$ & $ 7\,621$ & $17$ & $1.170$ & $3.949$ & $49$ \\
\hline
\hline
$3.0$  & Gauss    & $1024$ & $24\,352$ & $14$ & $2.340$ & $3.330$ & $16$ \\
$3.0$  & Gauss    & $2048$ & $12\,956$ & $14$ & $2.340$ & $3.330$ & $16$ \\
$3.0$  & Gauss    & $4096$ & $13\,039$ & $14$ & $2.340$ & $3.330$ & $16$ \\
$3.0$  & Gauss    & $8192$ & $ 7\,987$ & $14$ & $2.340$ & $3.330$ & $16$ \\
\hline
$3.5$  & Gauss    & $1024$ & $11\,697$ & $14$ & $1.755$ & $3.260$ & $25$ \\
$3.5$  & Gauss    & $2048$ & $19\,776$ & $14$ & $1.755$ & $3.260$ & $25$ \\
$3.5$  & Gauss    & $4096$ & $ 9\,367$ & $15$ & $1.755$ & $3.260$ & $25$ \\
$3.5$  & Gauss    & $8192$ & $10\,192$ & $16$ & $1.755$ & $3.260$ & $25$ \\
\hline
$4.5$  & Gauss    & $1024$ & $13\,673$ & $16$ & $1.027$ & $2.410$ & $27$ \\
$4.5$  & Gauss    & $2048$ & $10\,224$ & $16$ & $1.027$ & $2.410$ & $27$ \\
$4.5$  & Gauss    & $4096$ & $ 4\,656$ & $16$ & $1.027$ & $2.410$ & $27$ \\
$4.5$  & Gauss    & $8192$ & $ 8\,618$ & $16$ & $1.027$ & $2.410$ & $27$ \\
\hline
\hline
\end{tabular*}
\end{table}

\section{Numerical results}
\label{sec:mcres}

We first study Gaussian disorder where we have a strong equilibration
test to ensure that the data are in thermal equilibrium.  Corrections to
scaling are very large for this model despite the large number of
samples studied.

\subsection{Gaussian disorder}
\label{sec:gauss}

\begin{figure}[tb]
\includegraphics[width=\columnwidth]{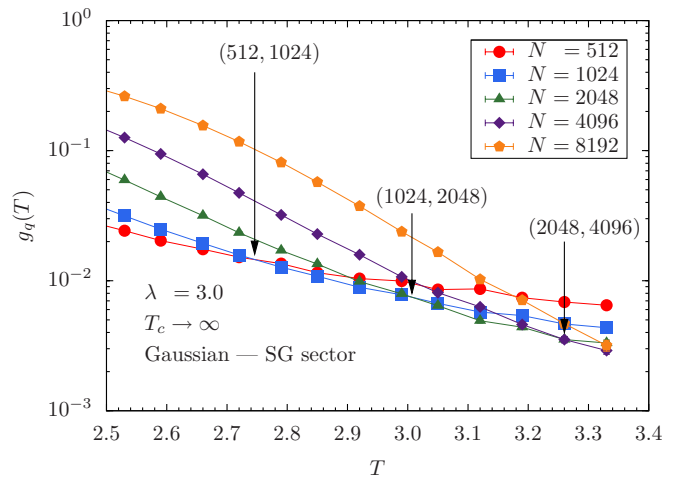}
\caption{(Color online)
Binder ratio $g_q$ for the SG sector and $\lambda = 3.0$ with Gaussian
disorder for different system sizes $N$. The crossings between
increasing system-size pairs diverge with increasing system size
suggesting that $T_c^{\rm SG} \to \infty$, in agreement with the
analytical predictions.
}
\label{fig:g-gauss-l3.000}
\end{figure}

\begin{figure}[tb]
\includegraphics[width=\columnwidth]{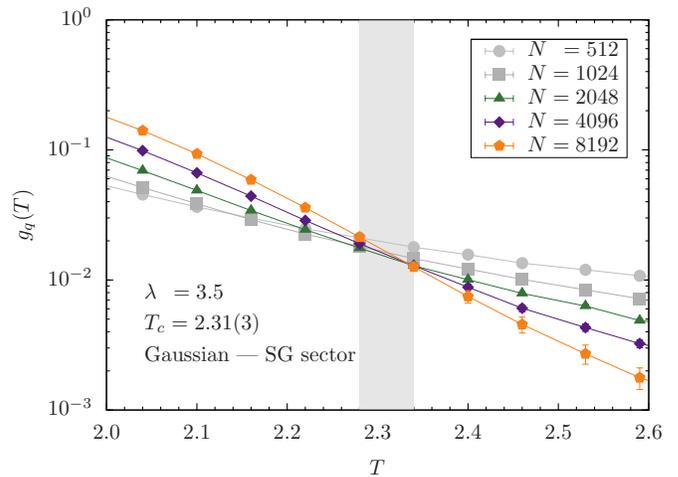}
\caption{(Color online)
Binder ratio $g_q$ for the SG sector and $\lambda = 3.5$ with Gaussian
disorder for different system sizes $N$. The data show strong
corrections to scaling. Using a finite-size scaling analysis we estimate
$T_c^{\rm SG} = 2.31(3)$. In this and all subsequent figures, the width
of the shaded region around the critical temperature corresponds to the
statistical uncertainty for the estimate of the critical temperature
using a combination of a finite-size scaling analysis and a bootstrap
method.
}
\label{fig:g-gauss-l3.500}
\end{figure}

\begin{figure}[tb]
\includegraphics[width=\columnwidth]{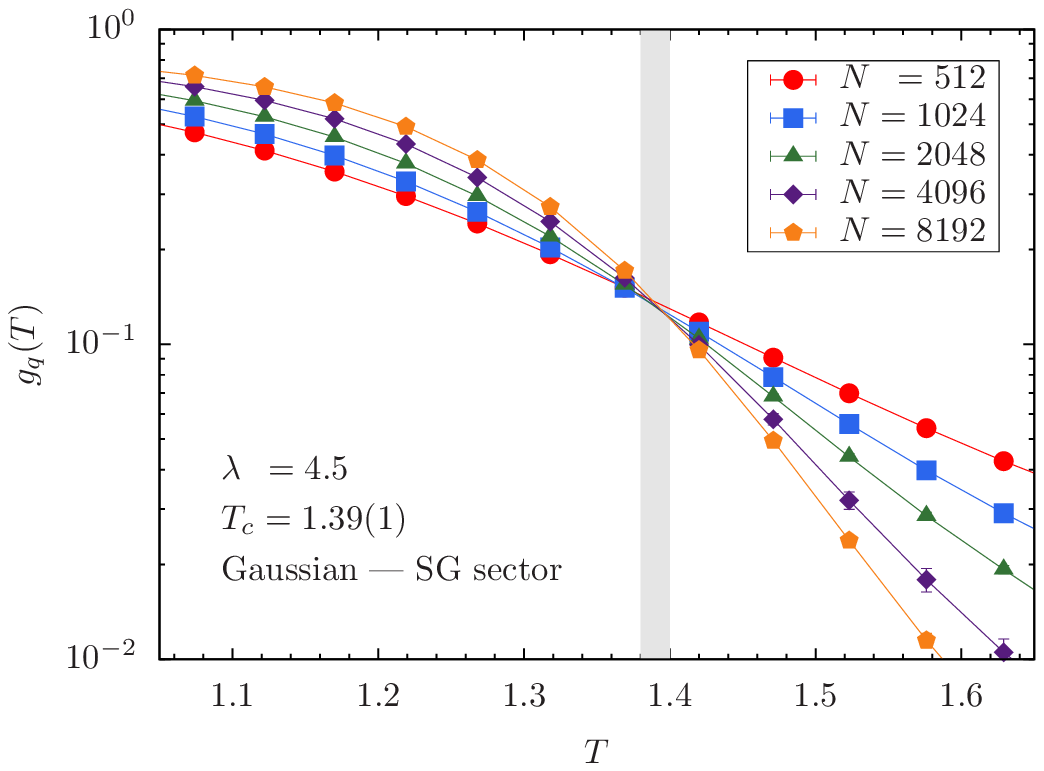}
\includegraphics[width=\columnwidth]{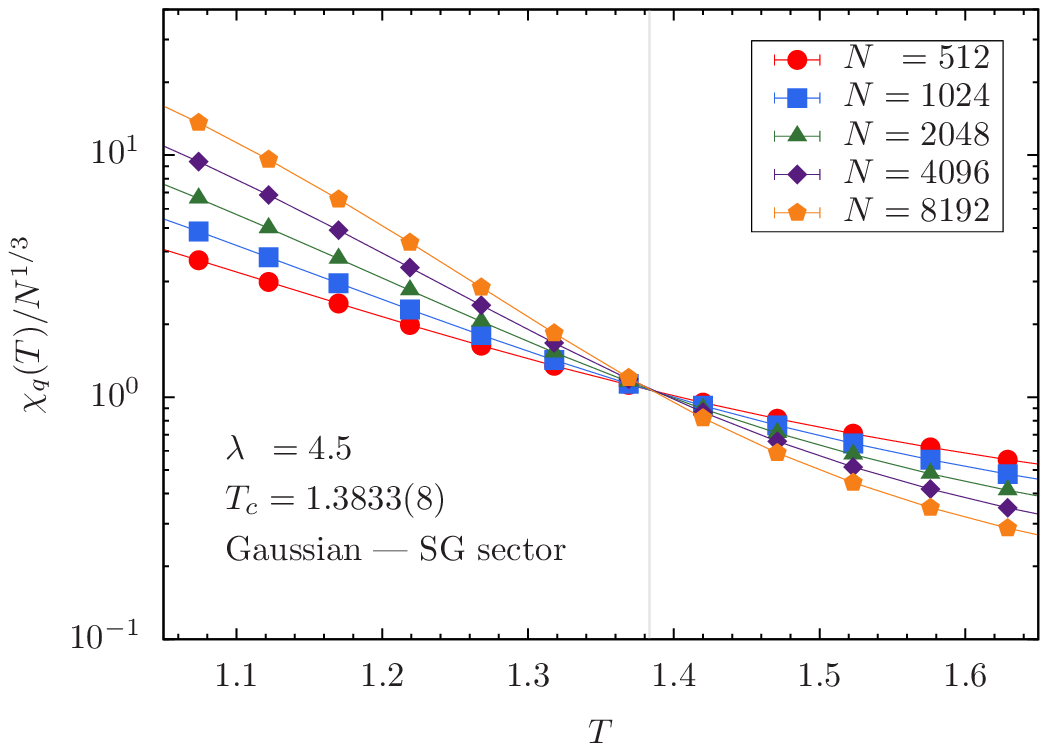}
\caption{(Color online)
Top: Binder ratio $g_q$ for the SG sector and $\lambda = 4.5$ with
Gaussian disorder for different system sizes $N$.  Corrections
to scaling are small and we estimate $T_c^{\rm SG} = 1.39(1)$.
Using Eq.~\eqref{eq:g_scale_mf} we obtain $T_c^{\rm SG} = 1.385(9)$,
which agrees within error bars with the previous estimate. Bottom:
Scaled spin-glass susceptibility $\chi_q/N^{1/3}$ with $\lambda = 4.5$
and Gaussian disorder for different system sizes $N$ as a function
of temperature. The data cross at a point (shaded area) and we obtain
$T_c^{\rm SG} = 1.3833(8)$.
}
\label{fig:g-gauss-l4.500}
\end{figure}

\begin{figure}[tb]
\includegraphics[width=\columnwidth]{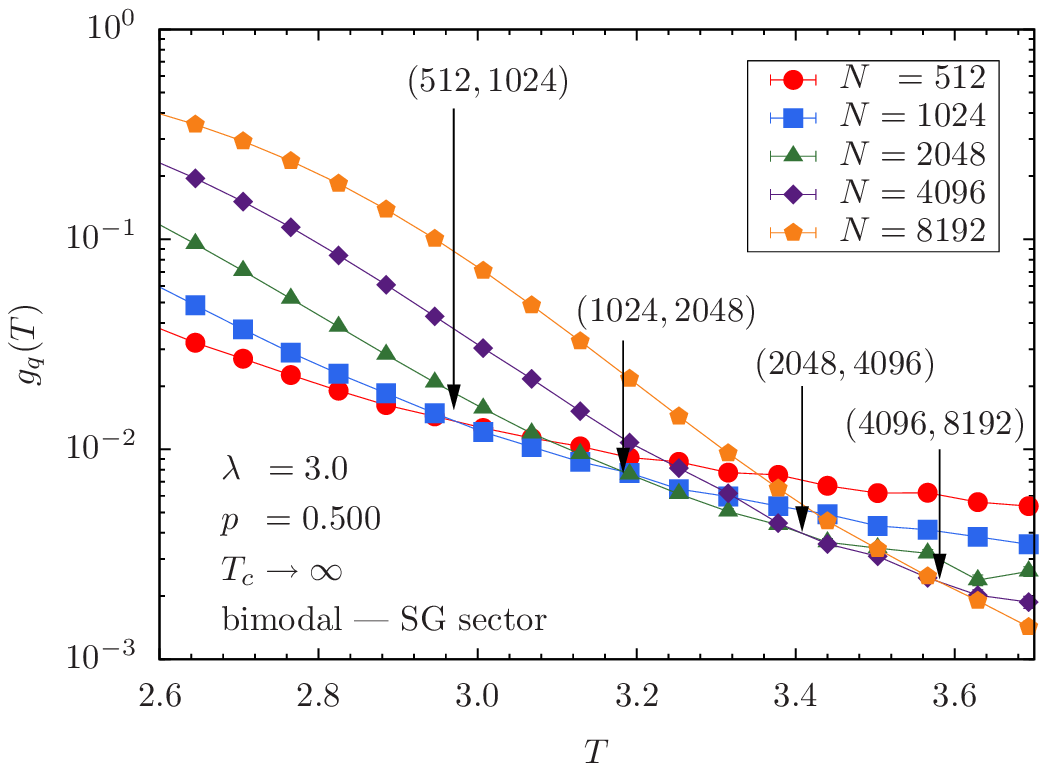}
\caption{(Color online)
Binder ratio $g_q$ for the SG sector and $\lambda = 3.0$ with bimodal
disorder for different system sizes $N$ and $p = 0.50$.  As for the
Gaussian case (Fig.~\ref{fig:g-gauss-l3.000}), the crossings between
increasing system-size pairs diverge with increasing system size
suggesting that $T_c^{\rm SG} \to \infty$, in agreement with the
analytical predictions.
}
\label{fig:g-pmj-l3.000-p0.500}
\end{figure}

Figure \ref{fig:g-gauss-l3.000} shows data for the Binder ratio $g_q$
for Gaussian disorder and $\lambda = 3.0$, right at the onset (see
Fig.~\ref{fig:phase_diagram}) where the critical temperature for the
spin-glass (SG) sector starts to diverge. The crossing temperatures
between lines for $N$/$2N$ pairs grow with the system size in agreement
with the analytic calculations. To prevent $T_c^{\rm SG}$ for the SG
sector to diverge, the bonds would have to be re-scaled. Furthermore,
there is no transition in the ferromagnetic sector (not shown), in
agreement with the analytical calculations.

In Fig.~\ref{fig:g-gauss-l3.500} we show data for $\lambda = 3.5$. In
agreement with the analytical predictions $T_c^{\rm SG}$ is finite,
albeit with huge corrections to scaling. We estimate $T_c^{\rm SG} =
2.31(3)$. Note that this estimate is computed via a finite-size scaling
of the data and only takes statistical fluctuations into account. We
have no control over finite-size corrections. A crude extrapolation
suggests that the critical temperature will likely be larger than the
quoted analytical value which we treat as a lower bound.  This is a generic 
problem for the estimates of the critical temperature made in the spin-glass
sector.  Again, there is no transition in the ferromagnetic sector.

\begin{figure}[tb]
\includegraphics[width=\columnwidth]{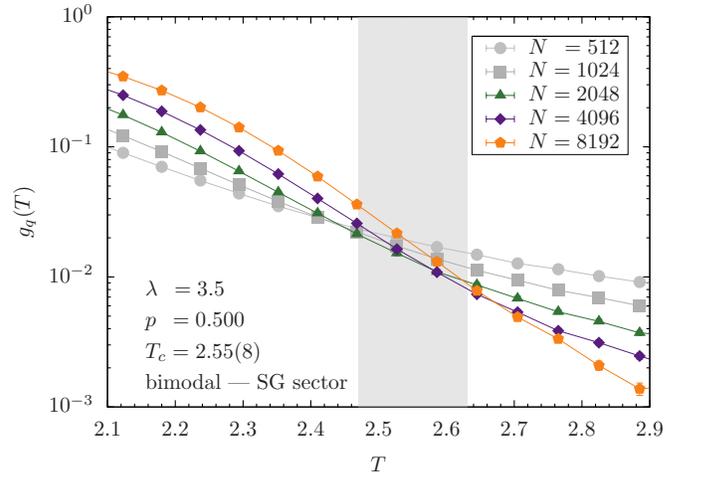}
\caption{(Color online)
Binder ratio $g_q$ for the SG sector and $\lambda = 3.5$ with bimodal
disorder for different system sizes $N$ and $p = 0.50$.  As for the
Gaussian case (Fig.~\ref{fig:g-gauss-l3.500}), the data show strong
corrections to scaling. Using a finite-size scaling analysis we estimate
$T_c^{\rm SG} = 2.55(8)$.
}
\label{fig:g-pmj-l3.500-p0.500}
\end{figure}

\begin{figure}[]

\includegraphics[width=\columnwidth]{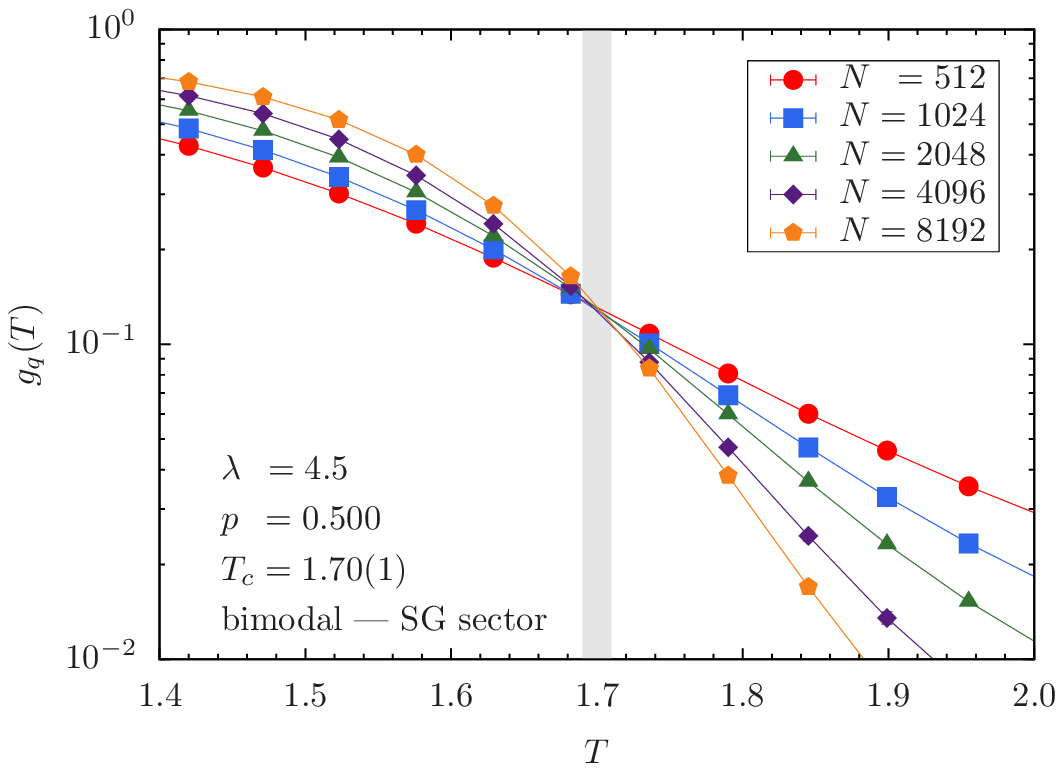}

\vspace*{0.1cm}

\includegraphics[width=\columnwidth]{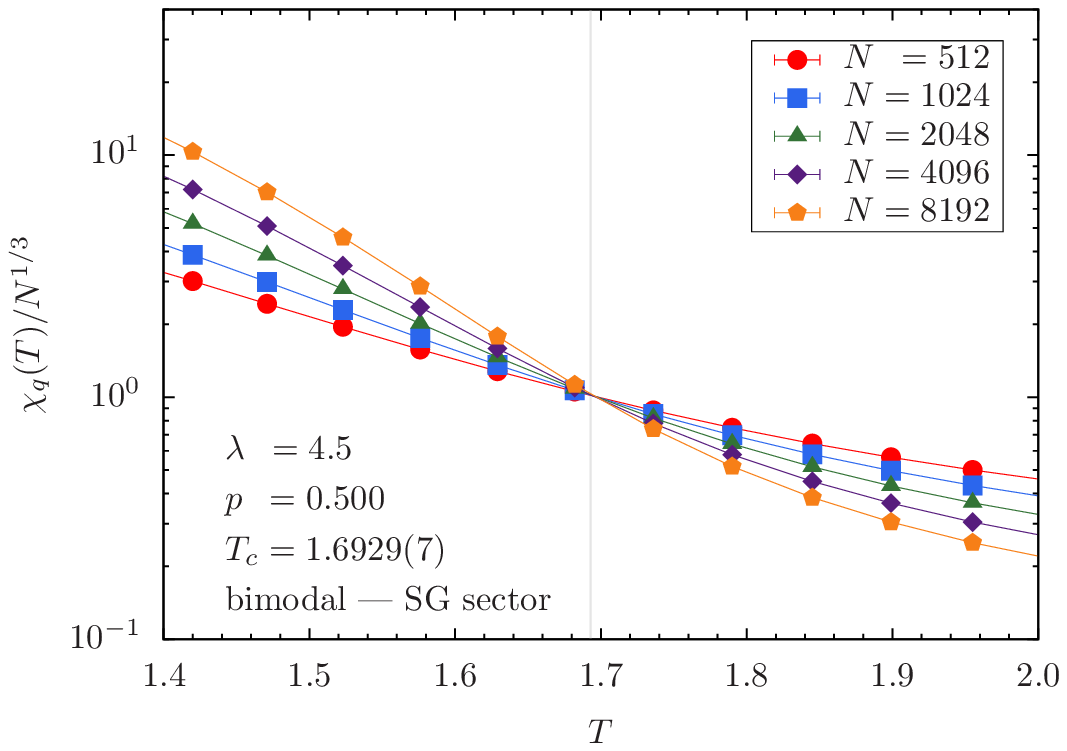}

\includegraphics[width=\columnwidth]{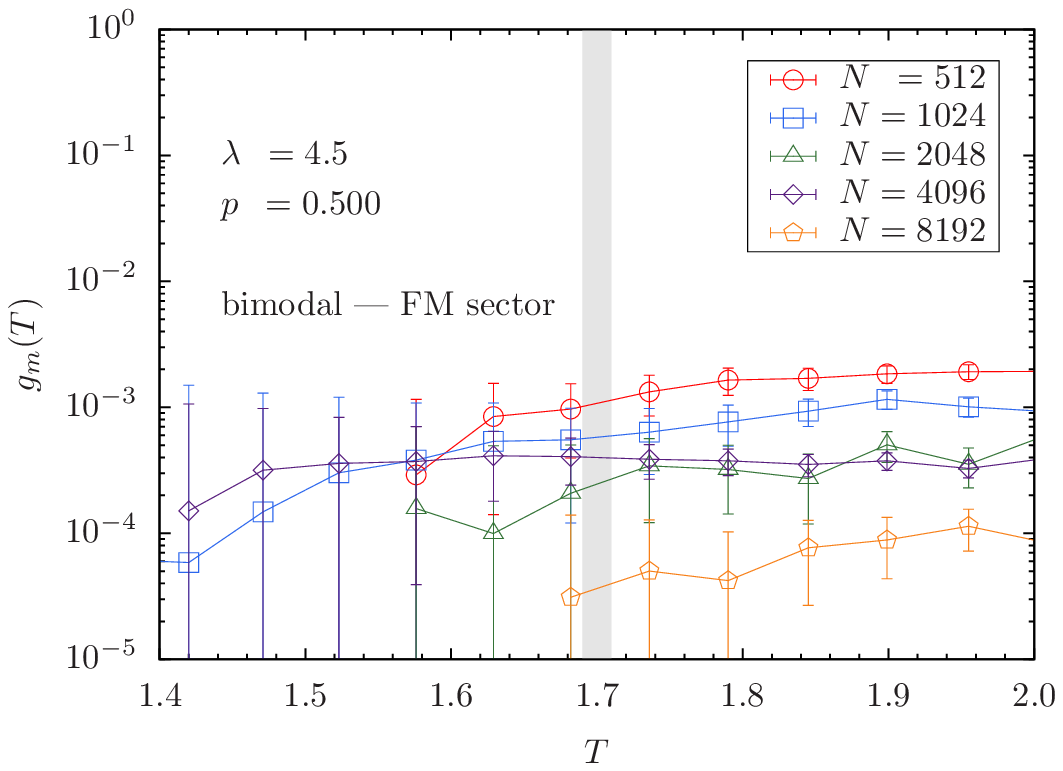}

\caption{(Color online)
Top panel: Binder ratio $g_q$ for the SG sector and $\lambda = 4.5$
with bimodal disorder ($p = 0.50$) for different system sizes $N$. We
estimate $T_c^{\rm SG} = 1.70(1)$.  Using Eq.~\eqref{eq:g_scale_mf} we
obtain $T_c^{\rm SG} = 1.695(8)$, which agrees within error bars with
the previous estimate.  Center panel: Scaled spin-glass susceptibility
$\chi_q/N^{1/3}$ as a function of temperature. The data cross at
a point (shaded area) and we obtain $T_c^{\rm SG} = 1.6929(7)$.
Bottom panel: Binder ratio $g_m$ for the FM sector. The shaded area
in the bottom panel corresponds to $T_c^{\rm SG} = 1.70 \pm 0.01$
(top panel). The data show no sign of ferromagnetic order: The data
do not cross and decrease for increasing system size $N$.
}
\label{fig:g-pmj-l4.500-p0.500}
\end{figure}

\begin{figure}[]

\includegraphics[width=\columnwidth]{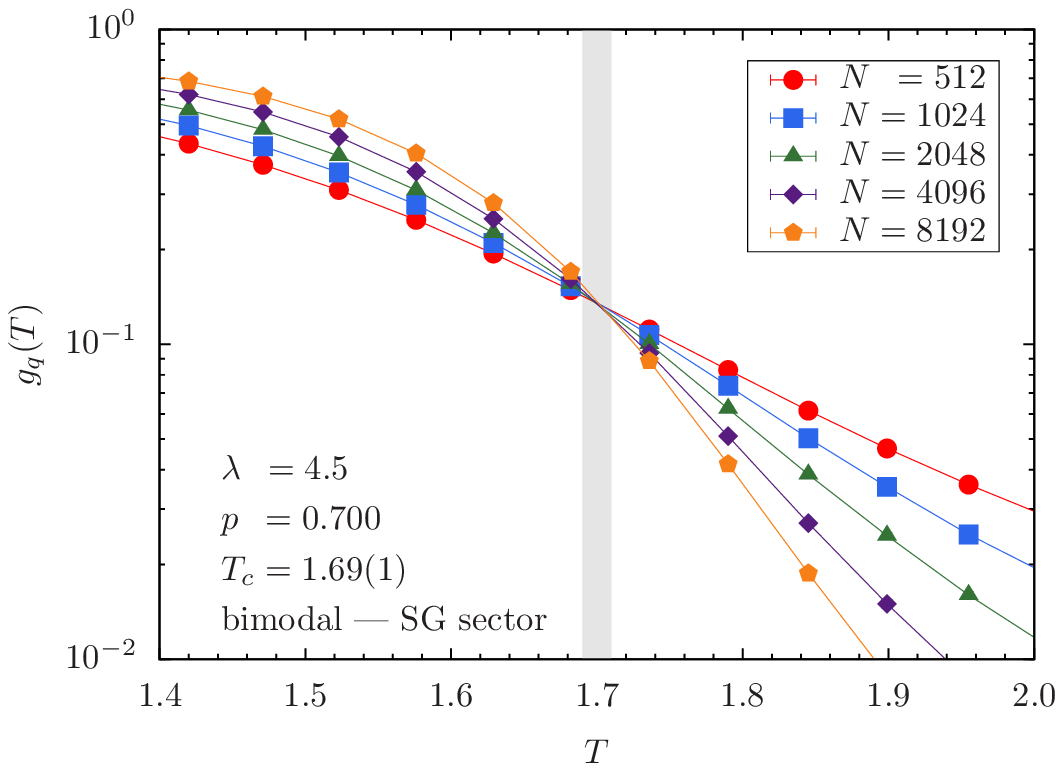}

\vspace*{0.1cm}

\includegraphics[width=\columnwidth]{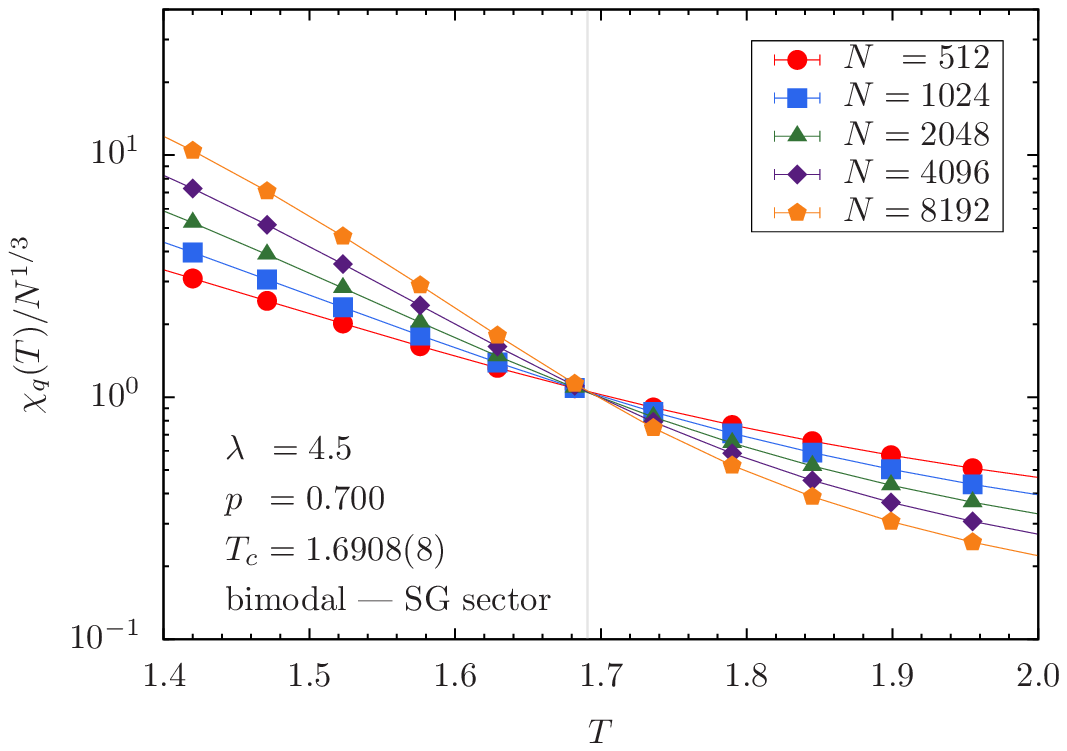}

\includegraphics[width=\columnwidth]{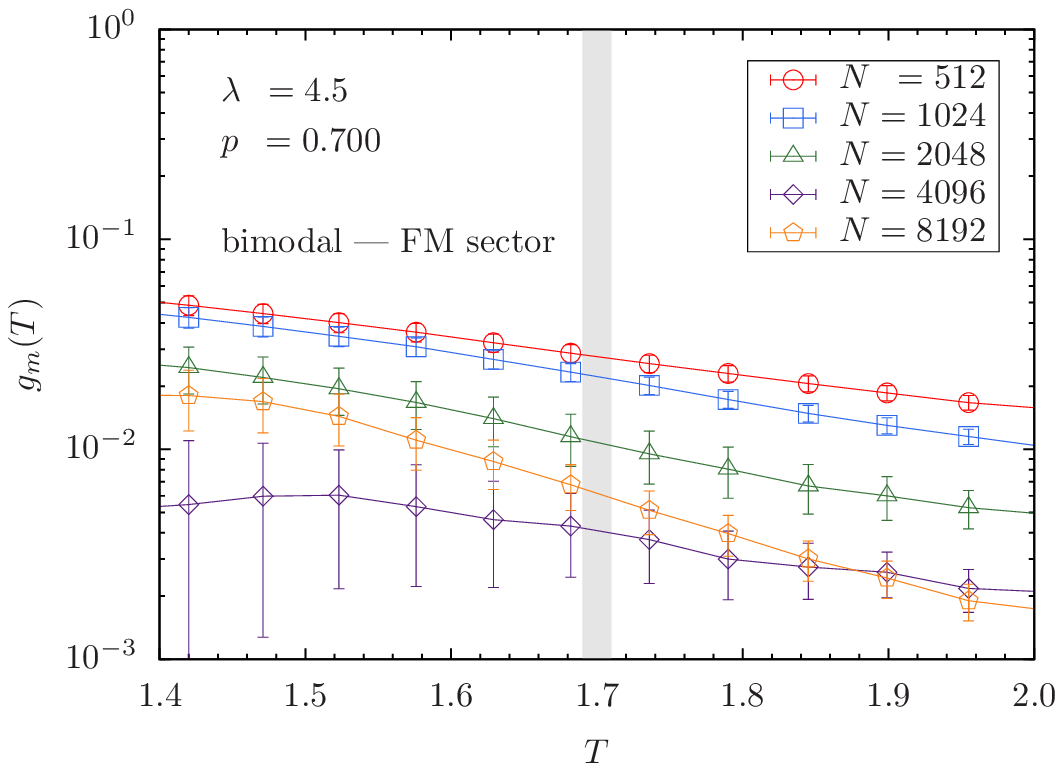}

\caption{(Color online)
Top panel: Binder ratio $g_q$ for the SG sector and $\lambda = 4.5$
with bimodal disorder ($p = 0.70$) for different system sizes $N$. We
estimate $T_c^{\rm SG} = 1.69(1)$.  Using Eq.~\eqref{eq:g_scale_mf} we
obtain $T_c^{\rm SG} = 1.693(9)$, which agrees within error bars with
the previous estimate.  Center panel: Scaled spin-glass susceptibility
$\chi_q/N^{1/3}$ as a function of temperature. The data cross at
a point (shaded area) and we obtain $T_c^{\rm SG} = 1.6908(8)$.
Bottom panel: Binder ratio $g_m$ for the FM sector. The shaded area
in the bottom panel corresponds to $T_c^{\rm SG} = 1.69 \pm 0.01$
(top panel). The data show no sign of ferromagnetic order: The data
do not cross and decrease for increasing system size $N$.
}
\label{fig:g-pmj-l4.500-p0.700}
\end{figure}

Finally, Fig.~\ref{fig:g-gauss-l4.500} shows data for $\lambda
= 4.5$. The top panel shows the Binder ratio as a function
of temperature.  The data cross at $T_c^{\rm SG} = 1.39(1)$.
Using Eq.~\eqref{eq:g_scale_mf} [i.e., fixing ``$\nu = 3$'' in
Eq.~\eqref{eq:g_scale}] we obtain $T_c^{\rm SG} = 1.385(9)$, which
agrees within error bars with the previous estimate.  This result
is verified by data on the scaled spin-glass susceptibility (bottom
panel of Fig.~\ref{fig:g-gauss-l4.500}) where the data also cross in
the same region. A finite-size scaling analysis of $\chi_q/N^{1/3}$
gives $T_c^{\rm SG} = 1.3833(8)$, a value of higher precision than
when using the Binder ratio.  For all values of $\lambda$ studied
where $T_c < \infty$ the estimates for the critical exponent $\nu$
are listed in Table \ref{tab:critparams}.  Note that for $\lambda > 4$,
``$\nu = 3$'' in the scaling form. However, we have allowed for the
value to fluctuate when estimating $T_c$ as well.

\subsection{Bimodal disorder}
\label{sec:pmj}

Figure \ref{fig:g-pmj-l3.000-p0.500} shows data for the Binder
ratio $g_q$ for bimodal disorder with $\lambda = 3.0$ and $p =
0.50$.  The chosen value of $\lambda$ is right at the onset (see
Fig.~\ref{fig:phase_diagram}) where the critical temperature for the
spin-glass SG starts to diverge. As for the Gaussian case
presented in Sec.~\ref{sec:gauss}, crossing points between lines
for $N$/$2N$ pairs grow with the system size in agreement with the
analytic calculations (i.e., $T_c^{\rm SG} \to \infty$).  Furthermore,
for the FM sector there is no transition (not shown).

In Fig.~\ref{fig:g-pmj-l3.500-p0.500} we show data for $\lambda = 3.5$
and $p = 0.50$. In agreement with the analytical predictions $T_c^{\rm
SG}$ is finite, albeit with large corrections to scaling. We estimate
$T_c^{\rm SG} = 2.55(8)$. Again, there is no ferromagnetic transition
(not shown).

We now study in detail $\lambda = 4.5$ for different concentrations
of ferromagnetic bonds $p$. Figure \ref{fig:g-pmj-l4.500-p0.500} shows
data for the Binder ratio for $\lambda = 4.5$ and $p = 0.50$ for both
SG and FM sectors. For the SG
sector (top panel), the data cross cleanly at $T_c^{\rm SG} = 1.70(1)$
[$1.695(8)$ when Eq.~\eqref{eq:g_scale_mf} is used].  The center
panel shows data for the scaled spin-glass susceptibility.  We obtain
$T_c^{\rm SG} = 1.6929(7)$, in agreement with the estimate from the
Binder ratio, albeit with higher precision.  However, for the FM sector
(bottom panel) the data strongly suggest that $T_c^{\rm FM}$ is not
defined, in agreement with the analytic predictions.  For $\lambda =
4.5$ the phase boundary between the FM and the SG sector lies somewhere
between $p = 0.7$ and $0.85$, see Fig.~\ref{fig:phase_diagram}.
Therefore, we expect that for $p = 0.7$ we only have SG order,
whereas for $p = 0.85$ the system orders ferromagnetically.

\begin{figure}[tb]
\includegraphics[width=\columnwidth]{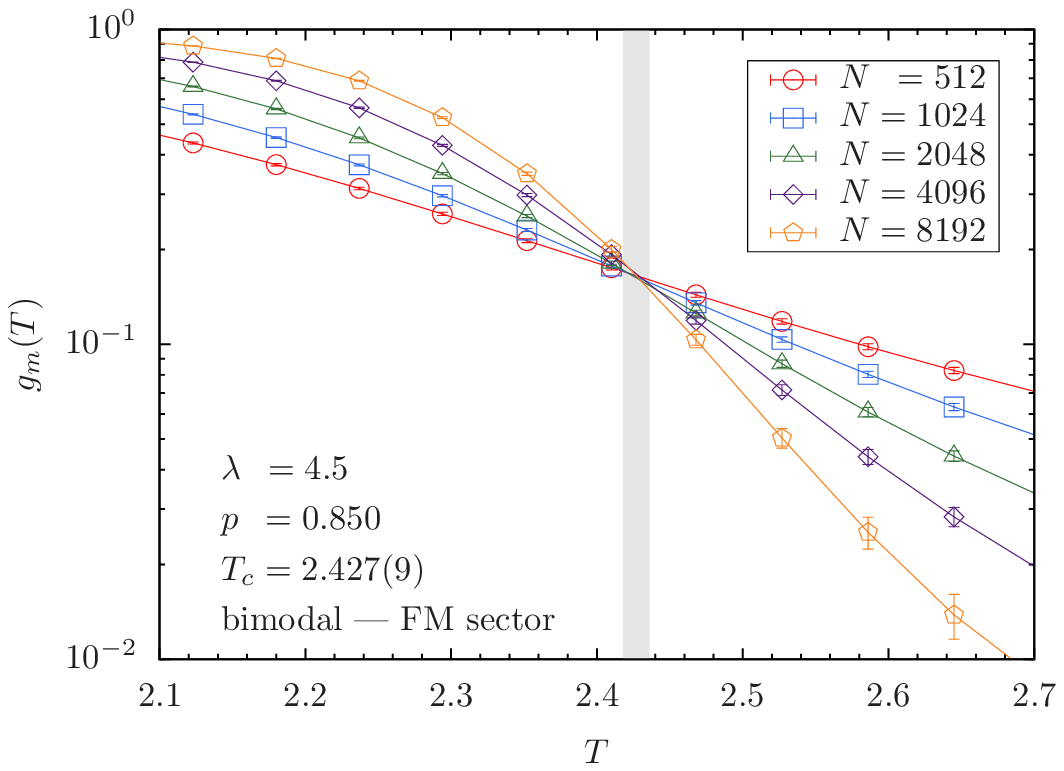}
\caption{(Color online)
Binder ratio $g_m$ for the FM sector and $p = 0.850$. The shaded area
corresponds to $T_c^{\rm FM} = 2.428(8)$.  In this case we are deep in
the FM sector, the SG sector shows no sign of a transition (not shown).
}
\label{fig:g-pmj-l4.500-p0.850}
\end{figure}

The case for $p = 0.70$ is shown in Fig.~\ref{fig:g-pmj-l4.500-p0.700}.
Like for $p = 0.50$, the data for the SG sector show a clear
transition, whereas the data for the FM sector suggest that there
is no transition.  Note that the estimates for $T_c^{\rm SG}$ agree
within error bars with the estimates for $p = 0.50$, suggesting that
the phase diagram does not depend on $p$ in the SG regime, in agreement
with the analytical calculations. For $p = 0.85$ there is no SG order
(not shown, in agreement with the analytical results).  Moreover,
$T_c^{\rm FM} = 2.428(8)$, see Fig.~\ref{fig:g-pmj-l4.500-p0.850}.

\subsection{Universality?}
\label{sec:uni}

To determine if two systems share the same universality class two
{\em independent} critical exponents need to be computed. However,
fluctuations in the data are large and therefore estimating the
critical exponent $\eta$ from a finite-size scaling analysis of
the susceptibility is difficult. Even worse, $\eta$ is not properly
defined for mean-field models and only $\alpha$, $\beta$, and $\gamma$
are available.

As shown above in Sec.~\ref{sec:analytical}, as well as the
work of Kim {\em et al.}~\cite{kim:05} and Dorogovtsev {\em et
al.}~\cite{dorogovtsev:08}, the critical exponents for spin glasses
and ferromagnets on scale-free graphs depend on the exponent
$\lambda$.  In particular, for $\lambda > 4$ the spin-glass
sector is predicted to share the same universality class as the
mean-field Sherrington-Kirkpatrick model.  For $\lambda < 4$ one
can show that for the critical exponent $\beta$ one finds $\beta =
\beta(\lambda)$. Similar predictions exist for the ferromagnetic
sector where mean-field behavior is recovered for $\lambda > 5$.
However, it remains to be determined if the type of disorder (e.g.,
Gaussian or bimodal) might change the critical exponents.

We compute the critical exponents in an unbiased fashion and with a
statistical error bar by letting $T_c$ and $y = 1/\nu$ be parameters.
Close to the transition temperature the scaling function can be
represented by a third-order polynomial. This is typically a very good
approximation. If the optimal values of the critical parameters $\nu$
and $T_c$ are chosen, then data for different system sizes should
collapse onto a single curve, the scaling function. Therefore,
by searching for the optimal fit to a third-order polynomial while
minimizing the chi-square of the fit with respect to the critical
parameters allows us to determine their optimal values.  Statistical
error bars to the optimal values are determined by a bootstrap
analysis. Note that these error bars take statistical fluctuations
into account but cannot properly account for systematic deviations
due to corrections to scaling \cite{comment:tc}.

\begin{table}
\caption{
Critical parameters $T_c$ \cite{comment:tc} and $\nu$, as well as the
Binder parameter at the transition temperature $g(T_c^{\rm SG})$, for
the spin glass (SG) and ferromagnetic (FM) sectors computed using a
finite-size scaling analysis of the data for $N \ge 2048$. Both $\nu$
and $g(T_c)$ are universal quantities. This means that if two systems
share the same universality class, the values of $\nu$ and $g(T_c)$
for both systems should agree. Note that for $\lambda > 4$ we have
used the scaling relation from Eq.~\eqref{eq:g_scale}. The obtained
exponent $y = 1/\nu$ is compatible with $y = 1/3$ as expected from
the mean-field solution. Columns for $\lambda = 4.5$ that state that
$\nu$ is ``fixed'' were computed using Eq.~\eqref{eq:g_scale_mf}.
Columns for $\lambda = 4.5$ that are marked with a $\dagger$ have
estimates of $T_c^{\rm SG}$ computed via a scaling of the susceptibility,
Eq.~\eqref{eq:chi_scale_mf}.
\label{tab:critparams}
}
\begin{tabular*}{\columnwidth}{@{\extracolsep{\fill}} l l r r r r r }
\hline
\hline
$\lambda$ & $p$ & $T_c^{\rm SG}$ & $\nu^{\rm SG}$ & $g(T_c^{\rm SG})$ & $T_c^{\rm FM}$ & $\nu^{\rm
FM}$ \\
\hline
$3.0$          & $0.500$ &  $\infty$   &       ---  &         --- &       ---  &       --- \\
\hline
$3.5$          & $0.500$ & $2.55(8)$   & $3.63(79)$ & $0.012(25)$ &       ---  &       --- \\
\hline
$4.5$          & $0.500$ & $1.70(1)$   & $3.43(51)$ & $0.129(12)$ &       ---  &       --- \\
$4.5$          & $0.500$ & $1.695(8)$  & fixed      & $0.134(9)$  &       ---  &       --- \\
$4.5^\dagger$  & $0.500$ & $1.6929(7)$ &        --- &        ---  &       ---  &       --- \\
\hline
$4.5$          & $0.700$ & $1.69(1)$   & $3.57(31)$ & $0.136(12)$ &       ---  &       --- \\
$4.5$          & $0.700$ & $1.693(9)$  & fixed      & $0.144(1)$  &       ---  &       --- \\
$4.5^\dagger$  & $0.700$ & $1.6908(8)$ &        --- &        ---  &       ---  &       --- \\
$4.5$          & $0.850$ &      ---    &       ---  &         --- & $2.428(8)$ & $2.70(9)$ \\
\hline
\hline
$3.0$          & Gauss   &  $\infty$   &       ---  &         --- &        --- &       --- \\
\hline
$3.5$          & Gauss   & $2.31(3)$   & $3.60(25)$ & $0.018(3)$  &        --- &       --- \\
\hline
$4.5$          & Gauss   & $1.39(1)$   & $3.53(58)$ & $0.132(12)$ &        --- &       --- \\
$4.5$          & Gauss   & $1.385(9)$  & fixed      & $0.138(9)$  &        --- &       --- \\
$4.5^\dagger$  & Gauss   & $1.3833(8)$ &        --- &        ---  &       --- &       --- \\

\hline
\hline
\end{tabular*}
\end{table}

To test for universality we estimate the critical exponent
$\nu$ and the value of the dimensionless Binder ratio at $T_c$
\cite{katzgraber:06} (see Table \ref{tab:critparams}).  Because
fluctuations in $\nu$ are very large and $\nu$ is not properly defined
for $\lambda > 4$ ($5$) in the spin-glass (ferromagnetic) sector we
compare in detail $g(T_c)$ in the spin-glass sector.  For $\lambda =
3.5$, $g(T_c^{\rm SG},\lambda = 3.5,{\rm Gauss}) = 0.018(3)$, which
agrees within error bars with $g(T_c^{\rm SG},\lambda = 3.5, p =
0.5) = 0.012(25)$.  For $\lambda = 4.5$, $g(T_c^{\rm SG},\lambda =
4.5,{\rm Gauss}) = 0.132(12)$, which agrees within error bars with
$g(T_c^{\rm SG},\lambda = 4.5,p = 0.5) = 0.129(12)$. In the bimodal
case, we find also that $g(T_c^{\rm SG},\lambda = 4.5,p = 0.5) =
0.129(12)$ agrees within error bars with $g(T_c^{\rm SG},\lambda =
4.5,p = 0.7) = 0.136(12)$, thus suggesting the the same universality
class might be shared below the spin-glass--to--ferromagnet phase
boundary for all values of $p$. The mean-field Viana-Bray model
\cite{viana:85} resembles a fixed-connectivity random graph and
is in the same universality class as the Sherrington-Kirkpatrick
model. Recent simulations \cite{wittmann:12} have shown that for the
mean-field universality class $g(T_c^{\rm SG}) \approx 0.126(46)$,
in agreement with our results for $\lambda > 4$.

Summarizing, our results suggest that for a given value of $\lambda$
the networks with both Gaussian and bimodal disorder share the {\em
same} universality class. In addition, for values of $\lambda >
4$ the spin-glass sector shares the same universality class as the
mean-field Sherrington-Kirkpatrick model.

\section{Conclusions}
\label{sec:concl}

We have studied Boolean (Ising) variables on a scale-free graph with
competing interactions. Our analytical and numerical results show that
for $\lambda \le 3$ the critical temperature diverges with the system
size. For larger values of $\lambda$ the system undergoes a
finite-temperature transition between a spin-glass and a paramagnetic
phase. The robustness of both the ferromagnetic and spin-glass phases
suggest that Boolean decision problems on scale-free networks are quite
stable to local (temperature) perturbations.  For the case with bimodal
disorder, we show that for a large enough fraction of ferromagnetic
bonds the system orders ferromagnetically at finite temperatures.
Finally, for a given value of $\lambda$ universal critical parameters for
both Gaussian and bimodal disorder agree, suggesting universal behavior.

Real networks typically have exponents $\lambda < 3$. It would be
interesting to study such a network in the future. Furthermore,
the effect of ``global biases'' (field terms) will also be studied.
Finally, opinion formation is an intrinsically nonequilibrium
process. For example, what are the temporal patterns of the agents
on the networks?  What are the effects of time-dependent interactions?

\begin{acknowledgments} 

We would like to thank Alexander K.~Hartmann, M.~Niemann, M.~Schechter,
M.~Wittmann and A.~P.~Young for fruitful discussions. We also thank
R.~S.~Andrist for assistance. H.G.K.~acknowledges
support from the SNF (Grant No.~PP002-114713) and the NSF (Grant
No.~DMR-1151387).  We would like to thank the Texas Advanced Computing
Center (TACC) at The University of Texas at Austin for providing HPC
resources (Ranger Sun Constellation Linux Cluster), ETH Zurich for
CPU time on the Brutus cluster, and Texas A\&M University for access
to their eos and lonestar clusters.

\end{acknowledgments}

\bibliography{refs,comments}

\begin{thebibliography}{53}
\expandafter\ifx\csname natexlab\endcsname\relax\def\natexlab#1{#1}\fi
\expandafter\ifx\csname bibnamefont\endcsname\relax
  \def\bibnamefont#1{#1}\fi
\expandafter\ifx\csname bibfnamefont\endcsname\relax
  \def\bibfnamefont#1{#1}\fi
\expandafter\ifx\csname citenamefont\endcsname\relax
  \def\citenamefont#1{#1}\fi
\expandafter\ifx\csname url\endcsname\relax
  \def\url#1{\texttt{#1}}\fi
\expandafter\ifx\csname urlprefix\endcsname\relax\def\urlprefix{URL }\fi
\providecommand{\bibinfo}[2]{#2}
\providecommand{\eprint}[2][]{\url{#2}}

\bibitem[{\citenamefont{Albert et~al.}(1999)\citenamefont{Albert, Jeong, and
  {Barab{\'a}si}}}]{albert:99}
\bibinfo{author}{\bibfnamefont{R.}~\bibnamefont{Albert}},
  \bibinfo{author}{\bibfnamefont{H.}~\bibnamefont{Jeong}}, \bibnamefont{and}
  \bibinfo{author}{\bibfnamefont{A.-L.} \bibnamefont{{Barab{\'a}si}}},
  \emph{\bibinfo{title}{{{Internet: Diameter of the World-Wide Web}}}},
  \bibinfo{journal}{Nature} \textbf{\bibinfo{volume}{401}},
  \bibinfo{pages}{130} (\bibinfo{year}{1999}).

\bibitem[{\citenamefont{Albert and Barab\'asi}(2002)}]{albert:02}
\bibinfo{author}{\bibfnamefont{R.}~\bibnamefont{Albert}} \bibnamefont{and}
  \bibinfo{author}{\bibfnamefont{A.-L.} \bibnamefont{Barab\'asi}},
  \emph{\bibinfo{title}{{Statistical mechanics of complex networks}}},
  \bibinfo{journal}{Rev. Mod. Phys.} \textbf{\bibinfo{volume}{74}},
  \bibinfo{pages}{47} (\bibinfo{year}{2002}).

\bibitem[{com({\natexlab{a}})}]{comment:facebook}
\bibinfo{note}{{http://facebook.com}}.

\bibitem[{com({\natexlab{b}})}]{comment:slashdot}
\bibinfo{note}{{http://slashdot.org}}.

\bibitem[{\citenamefont{Yeomans}(1992)}]{yeomans:92}
\bibinfo{author}{\bibfnamefont{J.~M.} \bibnamefont{Yeomans}},
  \emph{\bibinfo{title}{{Statistical Mechanics of Phase Transitions}}}
  (\bibinfo{publisher}{Oxford University Press}, \bibinfo{address}{Oxford},
  \bibinfo{year}{1992}).

\bibitem[{\citenamefont{Antall et~al.}(2006)\citenamefont{Antall, Krapivsky,
  and Redner}}]{antall:06}
\bibinfo{author}{\bibfnamefont{T.}~\bibnamefont{Antall}},
  \bibinfo{author}{\bibfnamefont{P.~L.} \bibnamefont{Krapivsky}},
  \bibnamefont{and} \bibinfo{author}{\bibfnamefont{S.}~\bibnamefont{Redner}},
  \emph{\bibinfo{title}{Social balance on networks: The dynamics of friendship
  and enmity}}, \bibinfo{journal}{Physica D} \textbf{\bibinfo{volume}{224}},
  \bibinfo{pages}{130} (\bibinfo{year}{2006}).

\bibitem[{\citenamefont{Leskovec et~al.}(2010)\citenamefont{Leskovec,
  Huttenlocher, and Kleinberg}}]{leskovec:10}
\bibinfo{author}{\bibfnamefont{J.}~\bibnamefont{Leskovec}},
  \bibinfo{author}{\bibfnamefont{D.}~\bibnamefont{Huttenlocher}},
  \bibnamefont{and}
  \bibinfo{author}{\bibfnamefont{J.}~\bibnamefont{Kleinberg}}, in
  \emph{\bibinfo{booktitle}{Proceedings of the 28th ACM Conference on Human
  Factors in Computing Systems}} (\bibinfo{year}{2010}), p.
  \bibinfo{pages}{1361}.

\bibitem[{\citenamefont{Bartolozzi et~al.}(2006)\citenamefont{Bartolozzi,
  Surungan, Leinweber, and Williams}}]{bartolozzi:06}
\bibinfo{author}{\bibfnamefont{M.}~\bibnamefont{Bartolozzi}},
  \bibinfo{author}{\bibfnamefont{T.}~\bibnamefont{Surungan}},
  \bibinfo{author}{\bibfnamefont{D.~B.} \bibnamefont{Leinweber}},
  \bibnamefont{and} \bibinfo{author}{\bibfnamefont{A.~G.}
  \bibnamefont{Williams}}, \emph{\bibinfo{title}{{{Spin-glass behavior of the
  antiferromagnetic Ising model on a scale-free network}}}},
  \bibinfo{journal}{Phys. Rev. B} \textbf{\bibinfo{volume}{73}},
  \bibinfo{pages}{224419} (\bibinfo{year}{2006}).

\bibitem[{\citenamefont{Herrero}(2009)}]{herrero:09}
\bibinfo{author}{\bibfnamefont{C.~P.} \bibnamefont{Herrero}},
  \emph{\bibinfo{title}{{{Antiferromagnetic Ising model in scale-free
  networks}}}}, \bibinfo{journal}{Eur. Phys. J. B}
  \textbf{\bibinfo{volume}{70}}, \bibinfo{pages}{435} (\bibinfo{year}{2009}).

\bibitem[{\citenamefont{Weigel and Johnston}(2007)}]{weigel:07}
\bibinfo{author}{\bibfnamefont{M.}~\bibnamefont{Weigel}} \bibnamefont{and}
  \bibinfo{author}{\bibfnamefont{D.}~\bibnamefont{Johnston}},
  \emph{\bibinfo{title}{{Frustration effects in antiferromagnets on planar
  random graphs}}}, \bibinfo{journal}{Phys. Rev. B}
  \textbf{\bibinfo{volume}{76}}, \bibinfo{pages}{054408}
  (\bibinfo{year}{2007}).

\bibitem[{\citenamefont{{Imry} and {Ma}}(1975)}]{imry:75}
\bibinfo{author}{\bibfnamefont{Y.}~\bibnamefont{{Imry}}} \bibnamefont{and}
  \bibinfo{author}{\bibfnamefont{S.-K.} \bibnamefont{{Ma}}},
  \emph{\bibinfo{title}{{{Random-Field Instability of the Ordered State of
  Continuous Symmetry}}}}, \bibinfo{journal}{Phys. Rev. Lett.}
  \textbf{\bibinfo{volume}{35}}, \bibinfo{pages}{1399} (\bibinfo{year}{1975}).

\bibitem[{\citenamefont{{Sethna} et~al.}(1993)\citenamefont{{Sethna}, {Dahmen},
  {Kartha}, {Krumhansl}, {Roberts}, and {Shore}}}]{sethna:93}
\bibinfo{author}{\bibfnamefont{J.~P.} \bibnamefont{{Sethna}}},
  \bibinfo{author}{\bibfnamefont{K.}~\bibnamefont{{Dahmen}}},
  \bibinfo{author}{\bibfnamefont{S.}~\bibnamefont{{Kartha}}},
  \bibinfo{author}{\bibfnamefont{J.~A.} \bibnamefont{{Krumhansl}}},
  \bibinfo{author}{\bibfnamefont{B.~W.} \bibnamefont{{Roberts}}},
  \bibnamefont{and} \bibinfo{author}{\bibfnamefont{J.~D.}
  \bibnamefont{{Shore}}}, \emph{\bibinfo{title}{Hysteresis and hierarchies:
  Dynamics of disorder-driven first-order phase transformations}},
  \bibinfo{journal}{Phys. Rev. Lett.} \textbf{\bibinfo{volume}{70}},
  \bibinfo{pages}{3347} (\bibinfo{year}{1993}).

\bibitem[{\citenamefont{Lee et~al.}(2006)\citenamefont{Lee, Jeong, and
  Noh}}]{lee:06a}
\bibinfo{author}{\bibfnamefont{S.~H.} \bibnamefont{Lee}},
  \bibinfo{author}{\bibfnamefont{H.}~\bibnamefont{Jeong}}, \bibnamefont{and}
  \bibinfo{author}{\bibfnamefont{J.~D.} \bibnamefont{Noh}},
  \emph{\bibinfo{title}{{{Random field Ising model on networks with
  inhomogeneous connections}}}}, \bibinfo{journal}{Phys. Rev. E}
  \textbf{\bibinfo{volume}{74}}, \bibinfo{pages}{031118}
  (\bibinfo{year}{2006}).

\bibitem[{\citenamefont{Mooij and Kappen}(2004)}]{mooij:04}
\bibinfo{author}{\bibfnamefont{J.~M.} \bibnamefont{Mooij}} \bibnamefont{and}
  \bibinfo{author}{\bibfnamefont{H.~J.} \bibnamefont{Kappen}},
  \emph{\bibinfo{title}{{{Spin-glass phase transitions on real-world graphs}}}}
  (\bibinfo{year}{2004}), \bibinfo{note}{(arXiv:cond-mat/0408378)}.

\bibitem[{\citenamefont{Kim et~al.}(2005)\citenamefont{Kim, Rodgers, Kahng, and
  Kim}}]{kim:05}
\bibinfo{author}{\bibfnamefont{D.-H.} \bibnamefont{Kim}},
  \bibinfo{author}{\bibfnamefont{G.~J.} \bibnamefont{Rodgers}},
  \bibinfo{author}{\bibfnamefont{B.}~\bibnamefont{Kahng}}, \bibnamefont{and}
  \bibinfo{author}{\bibfnamefont{D.}~\bibnamefont{Kim}},
  \emph{\bibinfo{title}{Spin-glass phase transition on scale-free networks}},
  \bibinfo{journal}{Phys. Rev. E} \textbf{\bibinfo{volume}{71}},
  \bibinfo{pages}{056115} (\bibinfo{year}{2005}).

\bibitem[{\citenamefont{Ferreira et~al.}(2010)\citenamefont{Ferreira, Mendes,
  and Ostilli}}]{ferreira:10}
\bibinfo{author}{\bibfnamefont{A.~L.} \bibnamefont{Ferreira}},
  \bibinfo{author}{\bibfnamefont{J.~F.~F.} \bibnamefont{Mendes}},
  \bibnamefont{and} \bibinfo{author}{\bibfnamefont{M.}~\bibnamefont{Ostilli}},
  \emph{\bibinfo{title}{{{First- and second-order phase transitions in Ising
  models on small-world networks: Simulations and comparison with an effective
  field theory}}}}, \bibinfo{journal}{Phys. Rev. E}
  \textbf{\bibinfo{volume}{82}}, \bibinfo{pages}{011141}
  (\bibinfo{year}{2010}).

\bibitem[{\citenamefont{Ostilli et~al.}(2011)\citenamefont{Ostilli, Ferreira,
  and Mendes}}]{ostilli:11}
\bibinfo{author}{\bibfnamefont{M.}~\bibnamefont{Ostilli}},
  \bibinfo{author}{\bibfnamefont{A.~L.} \bibnamefont{Ferreira}},
  \bibnamefont{and} \bibinfo{author}{\bibfnamefont{J.~F.~F.}
  \bibnamefont{Mendes}}, \emph{\bibinfo{title}{{{Critical behavior and
  correlations on scale-free small-world networks: Application to network
  design}}}}, \bibinfo{journal}{Phys. Rev. E} \textbf{\bibinfo{volume}{83}},
  \bibinfo{pages}{061149} (\bibinfo{year}{2011}).

\bibitem[{\citenamefont{Leone et~al.}(2002)\citenamefont{Leone, V\'{a}zquez,
  Vespignani, and Zecchina}}]{leone:02}
\bibinfo{author}{\bibfnamefont{M.}~\bibnamefont{Leone}},
  \bibinfo{author}{\bibfnamefont{A.}~\bibnamefont{V\'{a}zquez}},
  \bibinfo{author}{\bibfnamefont{A.}~\bibnamefont{Vespignani}},
  \bibnamefont{and} \bibinfo{author}{\bibfnamefont{R.}~\bibnamefont{Zecchina}},
  \emph{\bibinfo{title}{Ferromagnetic ordering in graphs with arbitrary degree
  distribution}}, \bibinfo{journal}{Eur. Phys. J. B}
  \textbf{\bibinfo{volume}{28}}, \bibinfo{pages}{191} (\bibinfo{year}{2002}).

\bibitem[{\citenamefont{Sherrington and Kirkpatrick}(1975)}]{sherrington:75}
\bibinfo{author}{\bibfnamefont{D.}~\bibnamefont{Sherrington}} \bibnamefont{and}
  \bibinfo{author}{\bibfnamefont{S.}~\bibnamefont{Kirkpatrick}},
  \emph{\bibinfo{title}{Solvable model of a spin glass}},
  \bibinfo{journal}{Phys. Rev. Lett.} \textbf{\bibinfo{volume}{35}},
  \bibinfo{pages}{1792} (\bibinfo{year}{1975}).

\bibitem[{com({\natexlab{c}})}]{comment:fm}
\bibinfo{note}{Note that similar behavior is found in the ferromagnetic sector
  where the change to the mean-field Ising universality class occurs at
  $\lambda = 5$ \cite{dorogovtsev:08}.}

\bibitem[{\citenamefont{Edwards and Anderson}(1975)}]{edwards:75}
\bibinfo{author}{\bibfnamefont{S.~F.} \bibnamefont{Edwards}} \bibnamefont{and}
  \bibinfo{author}{\bibfnamefont{P.~W.} \bibnamefont{Anderson}},
  \emph{\bibinfo{title}{Theory of spin glasses}}, \bibinfo{journal}{J. Phys. F:
  Met. Phys.} \textbf{\bibinfo{volume}{5}}, \bibinfo{pages}{965}
  (\bibinfo{year}{1975}).

\bibitem[{\citenamefont{Binder and Young}(1986)}]{binder:86}
\bibinfo{author}{\bibfnamefont{K.}~\bibnamefont{Binder}} \bibnamefont{and}
  \bibinfo{author}{\bibfnamefont{A.~P.} \bibnamefont{Young}},
  \emph{\bibinfo{title}{Spin glasses: Experimental facts, theoretical concepts
  and open questions}}, \bibinfo{journal}{Rev. Mod. Phys.}
  \textbf{\bibinfo{volume}{58}}, \bibinfo{pages}{801} (\bibinfo{year}{1986}).

\bibitem[{\citenamefont{Barabasi and Albert}(1999)}]{barabasi:99}
\bibinfo{author}{\bibfnamefont{A.~L.} \bibnamefont{Barabasi}} \bibnamefont{and}
  \bibinfo{author}{\bibfnamefont{R.}~\bibnamefont{Albert}},
  \bibinfo{journal}{Science} \textbf{\bibinfo{volume}{286}},
  \bibinfo{pages}{509} (\bibinfo{year}{1999}).

\bibitem[{\citenamefont{Krapivsky et~al.}(2000)\citenamefont{Krapivsky, Redner,
  and Leyvraz}}]{krapivsky:00}
\bibinfo{author}{\bibfnamefont{P.~L.} \bibnamefont{Krapivsky}},
  \bibinfo{author}{\bibfnamefont{S.}~\bibnamefont{Redner}}, \bibnamefont{and}
  \bibinfo{author}{\bibfnamefont{F.}~\bibnamefont{Leyvraz}},
  \emph{\bibinfo{title}{Connectivity of growing random networks}},
  \bibinfo{journal}{Phys. Rev. Lett.} \textbf{\bibinfo{volume}{85}},
  \bibinfo{pages}{4629} (\bibinfo{year}{2000}).

\bibitem[{\citenamefont{Krapivsky and Redner}(2001)}]{krapivsky:01}
\bibinfo{author}{\bibfnamefont{P.~L.} \bibnamefont{Krapivsky}}
  \bibnamefont{and} \bibinfo{author}{\bibfnamefont{S.}~\bibnamefont{Redner}},
  \emph{\bibinfo{title}{Organization of growing random networks}},
  \bibinfo{journal}{Phys. Rev. E} \textbf{\bibinfo{volume}{63}},
  \bibinfo{pages}{066123} (\bibinfo{year}{2001}).

\bibitem[{\citenamefont{Bender and Canfield}(1978)}]{bender:78}
\bibinfo{author}{\bibfnamefont{E.~A.} \bibnamefont{Bender}} \bibnamefont{and}
  \bibinfo{author}{\bibfnamefont{E.~R.} \bibnamefont{Canfield}},
  \emph{\bibinfo{title}{{The asymptotic number of labeled graphs with given
  degree sequences}}}, \bibinfo{journal}{J. Comb. Theory A}
  \textbf{\bibinfo{volume}{24}}, \bibinfo{pages}{296} (\bibinfo{year}{1978}).

\bibitem[{\citenamefont{Newman}(2003)}]{newman:03a}
\bibinfo{author}{\bibfnamefont{M.~E.~J.} \bibnamefont{Newman}}, in
  \emph{\bibinfo{booktitle}{Handbook of Graphs and Networks}}, edited by
  \bibinfo{editor}{\bibfnamefont{S.}~\bibnamefont{Bornholdt}} \bibnamefont{and}
  \bibinfo{editor}{\bibfnamefont{H.~G.} \bibnamefont{Schuster}}
  (\bibinfo{publisher}{Wiley-VCH}, \bibinfo{address}{Berlin},
  \bibinfo{year}{2003}).

\bibitem[{\citenamefont{Catanzaro et~al.}(2005)\citenamefont{Catanzaro,
  Bogu\~{n}\'{a}, and Pastor-Satorras}}]{catanzaro:05}
\bibinfo{author}{\bibfnamefont{M.}~\bibnamefont{Catanzaro}},
  \bibinfo{author}{\bibfnamefont{M.}~\bibnamefont{Bogu\~{n}\'{a}}},
  \bibnamefont{and}
  \bibinfo{author}{\bibfnamefont{R.}~\bibnamefont{Pastor-Satorras}},
  \emph{\bibinfo{title}{Generation of uncorrelated random scale-free
  networks}}, \bibinfo{journal}{Phys. Rev. E} \textbf{\bibinfo{volume}{71}},
  \bibinfo{pages}{027103} (\bibinfo{year}{2005}).

\bibitem[{\citenamefont{Klein-Hennig and Hartmann}(2011)}]{klein-hennig:11}
\bibinfo{author}{\bibfnamefont{H.}~\bibnamefont{Klein-Hennig}}
  \bibnamefont{and} \bibinfo{author}{\bibfnamefont{A.~K.}
  \bibnamefont{Hartmann}}, \emph{\bibinfo{title}{Bias in generation of random
  graphs}} (\bibinfo{year}{2011}), \bibinfo{note}{(arxiv:cond-mat/1107.5734)}.

\bibitem[{com({\natexlab{d}})}]{comment:akh}
\bibinfo{note}{Note that the method we use to generate the scale-free graphs
  also suffers from the bias recently characterized by the authors of
  Ref.~\cite{klein-hennig:11}. However, for $\lambda \ge 3$, which are the
  values we are interested in, this has no effect on the data.}

\bibitem[{\citenamefont{Burda and Krzywicki}(2003)}]{burda:03}
\bibinfo{author}{\bibfnamefont{Z.}~\bibnamefont{Burda}} \bibnamefont{and}
  \bibinfo{author}{\bibfnamefont{A.}~\bibnamefont{Krzywicki}},
  \emph{\bibinfo{title}{{Uncorrelated random networks}}},
  \bibinfo{journal}{Phys. Rev. E} \textbf{\bibinfo{volume}{67}},
  \bibinfo{pages}{046118} (\bibinfo{year}{2003}).

\bibitem[{\citenamefont{Bogu\~{n}\'a et~al.}(2004)\citenamefont{Bogu\~{n}\'a,
  Pastor-Satorras, and Vespignani}}]{boguna:04}
\bibinfo{author}{\bibfnamefont{M.}~\bibnamefont{Bogu\~{n}\'a}},
  \bibinfo{author}{\bibfnamefont{R.}~\bibnamefont{Pastor-Satorras}},
  \bibnamefont{and}
  \bibinfo{author}{\bibfnamefont{A.}~\bibnamefont{Vespignani}},
  \emph{\bibinfo{title}{{Cut-offs and finite size effects in scale-free
  networks}}}, \bibinfo{journal}{Eur. Phys. J. B}
  \textbf{\bibinfo{volume}{38}}, \bibinfo{pages}{205} (\bibinfo{year}{2004}).

\bibitem[{\citenamefont{Wemmenhove et~al.}(2005)\citenamefont{Wemmenhove,
  Nikoletopoulos, and Hatchett}}]{wemmenhove:05}
\bibinfo{author}{\bibfnamefont{B.}~\bibnamefont{Wemmenhove}},
  \bibinfo{author}{\bibfnamefont{T.}~\bibnamefont{Nikoletopoulos}},
  \bibnamefont{and} \bibinfo{author}{\bibfnamefont{J.~P.~L.}
  \bibnamefont{Hatchett}}, \emph{\bibinfo{title}{{{Replica symmetry breaking in
  the 'small world' spin glass}}}}, \bibinfo{journal}{J. Stat. Mech.}
  \textbf{\bibinfo{volume}{\normalfont{P11007}}} (\bibinfo{year}{2005}).

\bibitem[{\citenamefont{{M{\'e}zard} and {Parisi}}(2001)}]{mezard:01}
\bibinfo{author}{\bibfnamefont{M.}~\bibnamefont{{M{\'e}zard}}}
  \bibnamefont{and} \bibinfo{author}{\bibfnamefont{G.}~\bibnamefont{{Parisi}}},
  \emph{\bibinfo{title}{{{The Bethe lattice spin glass revisited}}}},
  \bibinfo{journal}{Eur. Phys. J. B} \textbf{\bibinfo{volume}{20}},
  \bibinfo{pages}{217} (\bibinfo{year}{2001}).

\bibitem[{\citenamefont{Viana and Bray}(1985)}]{viana:85}
\bibinfo{author}{\bibfnamefont{L.}~\bibnamefont{Viana}} \bibnamefont{and}
  \bibinfo{author}{\bibfnamefont{A.~J.} \bibnamefont{Bray}},
  \emph{\bibinfo{title}{Phase diagrams for dilute spin glasses}},
  \bibinfo{journal}{J. Phys. C} \textbf{\bibinfo{volume}{18}},
  \bibinfo{pages}{3037} (\bibinfo{year}{1985}).

\bibitem[{\citenamefont{Monasson}(1998)}]{monasson:98}
\bibinfo{author}{\bibfnamefont{R.}~\bibnamefont{Monasson}},
  \emph{\bibinfo{title}{{{Optimization problems and replica symmetry breaking
  in finite connectivity spin glasses}}}}, \bibinfo{journal}{J. Phys. A}
  \textbf{\bibinfo{volume}{31}}, \bibinfo{pages}{513} (\bibinfo{year}{1998}).

\bibitem[{\citenamefont{de~Almeida and Thouless}(1978)}]{almeida:78}
\bibinfo{author}{\bibfnamefont{J.~R.~L.} \bibnamefont{de~Almeida}}
  \bibnamefont{and} \bibinfo{author}{\bibfnamefont{D.~J.}
  \bibnamefont{Thouless}}, \emph{\bibinfo{title}{Stability of the
  {S}herrington-{K}irkpatrick solution of a spin glass model}},
  \bibinfo{journal}{J. Phys. A} \textbf{\bibinfo{volume}{11}},
  \bibinfo{pages}{983} (\bibinfo{year}{1978}).

\bibitem[{\citenamefont{Cizeau and Bouchaud}(1993)}]{cizeau:93}
\bibinfo{author}{\bibfnamefont{P.}~\bibnamefont{Cizeau}} \bibnamefont{and}
  \bibinfo{author}{\bibfnamefont{J.~P.} \bibnamefont{Bouchaud}},
  \emph{\bibinfo{title}{{{Mean field theory of dilute spin-glasses with
  power-law interactions}}}}, \bibinfo{journal}{J. Phys. A}
  \textbf{\bibinfo{volume}{26}}, \bibinfo{pages}{L187} (\bibinfo{year}{1993}).

\bibitem[{\citenamefont{Neri et~al.}(2010)\citenamefont{Neri, Metz, and
  {Boll{\'e}}}}]{neri:10}
\bibinfo{author}{\bibfnamefont{I.}~\bibnamefont{Neri}},
  \bibinfo{author}{\bibfnamefont{F.~L.} \bibnamefont{Metz}}, \bibnamefont{and}
  \bibinfo{author}{\bibfnamefont{D.}~\bibnamefont{{Boll{\'e}}}},
  \emph{\bibinfo{title}{{{The phase diagram of L{\'e}vy spin glasses}}}},
  \bibinfo{journal}{J. Stat. Mech.}
  \textbf{\bibinfo{volume}{\normalfont{P01010}}} (\bibinfo{year}{2010}).

\bibitem[{\citenamefont{Janzen et~al.}(2010)\citenamefont{Janzen, Engel, and
  {M{\'e}zard}}}]{janzen:10a}
\bibinfo{author}{\bibfnamefont{K.}~\bibnamefont{Janzen}},
  \bibinfo{author}{\bibfnamefont{A.}~\bibnamefont{Engel}}, \bibnamefont{and}
  \bibinfo{author}{\bibfnamefont{M.}~\bibnamefont{{M{\'e}zard}}},
  \emph{\bibinfo{title}{{Thermodynamics of the L{\'e}vy spin glass}}},
  \bibinfo{journal}{Phys. Rev. E} \textbf{\bibinfo{volume}{82}},
  \bibinfo{pages}{021127} (\bibinfo{year}{2010}).

\bibitem[{\citenamefont{Binder}(1981)}]{binder:81}
\bibinfo{author}{\bibfnamefont{K.}~\bibnamefont{Binder}},
  \emph{\bibinfo{title}{Critical properties from {M}onte {C}arlo coarse
  graining and renormalization}}, \bibinfo{journal}{Phys. Rev. Lett.}
  \textbf{\bibinfo{volume}{47}}, \bibinfo{pages}{693} (\bibinfo{year}{1981}).

\bibitem[{\citenamefont{Larson et~al.}(2010)\citenamefont{Larson, Katzgraber,
  Moore, and Young}}]{larson:10}
\bibinfo{author}{\bibfnamefont{D.}~\bibnamefont{Larson}},
  \bibinfo{author}{\bibfnamefont{H.~G.} \bibnamefont{Katzgraber}},
  \bibinfo{author}{\bibfnamefont{M.~A.} \bibnamefont{Moore}}, \bibnamefont{and}
  \bibinfo{author}{\bibfnamefont{A.~P.} \bibnamefont{Young}},
  \emph{\bibinfo{title}{Numerical studies of a one-dimensional 3-spin
  spin-glass model with long-range interactions}}, \bibinfo{journal}{Phys. Rev.
  B} \textbf{\bibinfo{volume}{81}}, \bibinfo{pages}{064415}
  (\bibinfo{year}{2010}).

\bibitem[{\citenamefont{Cooper et~al.}(1982)\citenamefont{Cooper, Freedman, and
  Preston}}]{cooper:82}
\bibinfo{author}{\bibfnamefont{F.}~\bibnamefont{Cooper}},
  \bibinfo{author}{\bibfnamefont{B.}~\bibnamefont{Freedman}}, \bibnamefont{and}
  \bibinfo{author}{\bibfnamefont{D.}~\bibnamefont{Preston}},
  \emph{\bibinfo{title}{Solving $\phi^4_{1,2}$ theory with {M}onte {C}arlo}},
  \bibinfo{journal}{Nucl. Phys. B} \textbf{\bibinfo{volume}{210}},
  \bibinfo{pages}{210} (\bibinfo{year}{1982}).

\bibitem[{\citenamefont{Palassini and Caracciolo}(1999)}]{palassini:99b}
\bibinfo{author}{\bibfnamefont{M.}~\bibnamefont{Palassini}} \bibnamefont{and}
  \bibinfo{author}{\bibfnamefont{S.}~\bibnamefont{Caracciolo}},
  \emph{\bibinfo{title}{{U}niversal {F}inite-{S}ize {S}caling {F}unctions in
  the 3{D} {I}sing {S}pin {G}lass}}, \bibinfo{journal}{Phys. Rev. Lett.}
  \textbf{\bibinfo{volume}{82}}, \bibinfo{pages}{5128} (\bibinfo{year}{1999}).

\bibitem[{\citenamefont{Ballesteros et~al.}(2000)\citenamefont{Ballesteros,
  Cruz, Fernandez, Martin-Mayor, Pech, Ruiz-Lorenzo, Tarancon, Tellez, Ullod,
  and Ungil}}]{ballesteros:00}
\bibinfo{author}{\bibfnamefont{H.~G.} \bibnamefont{Ballesteros}},
  \bibinfo{author}{\bibfnamefont{A.}~\bibnamefont{Cruz}},
  \bibinfo{author}{\bibfnamefont{L.~A.} \bibnamefont{Fernandez}},
  \bibinfo{author}{\bibfnamefont{V.}~\bibnamefont{Martin-Mayor}},
  \bibinfo{author}{\bibfnamefont{J.}~\bibnamefont{Pech}},
  \bibinfo{author}{\bibfnamefont{J.~J.} \bibnamefont{Ruiz-Lorenzo}},
  \bibinfo{author}{\bibfnamefont{A.}~\bibnamefont{Tarancon}},
  \bibinfo{author}{\bibfnamefont{P.}~\bibnamefont{Tellez}},
  \bibinfo{author}{\bibfnamefont{C.~L.} \bibnamefont{Ullod}}, \bibnamefont{and}
  \bibinfo{author}{\bibfnamefont{C.}~\bibnamefont{Ungil}},
  \emph{\bibinfo{title}{Critical behavior of the three-dimensional {I}sing spin
  glass}}, \bibinfo{journal}{Phys. Rev. B} \textbf{\bibinfo{volume}{62}},
  \bibinfo{pages}{14237} (\bibinfo{year}{2000}).

\bibitem[{\citenamefont{{Mart{\'{\i}}n-Mayor}
  et~al.}(2002)\citenamefont{{Mart{\'{\i}}n-Mayor}, {Pelissetto}, and
  {Vicari}}}]{martin:02}
\bibinfo{author}{\bibfnamefont{V.}~\bibnamefont{{Mart{\'{\i}}n-Mayor}}},
  \bibinfo{author}{\bibfnamefont{A.}~\bibnamefont{{Pelissetto}}},
  \bibnamefont{and} \bibinfo{author}{\bibfnamefont{E.}~\bibnamefont{{Vicari}}},
  \emph{\bibinfo{title}{{Critical structure factor in Ising systems}}},
  \bibinfo{journal}{Phys. Rev. E} \textbf{\bibinfo{volume}{66}},
  \bibinfo{pages}{026112} (\bibinfo{year}{2002}).

\bibitem[{\citenamefont{Hukushima and Nemoto}(1996)}]{hukushima:96}
\bibinfo{author}{\bibfnamefont{K.}~\bibnamefont{Hukushima}} \bibnamefont{and}
  \bibinfo{author}{\bibfnamefont{K.}~\bibnamefont{Nemoto}},
  \emph{\bibinfo{title}{Exchange {M}onte {C}arlo method and application to spin
  glass simulations}}, \bibinfo{journal}{J. Phys. Soc. Jpn.}
  \textbf{\bibinfo{volume}{65}}, \bibinfo{pages}{1604} (\bibinfo{year}{1996}).

\bibitem[{\citenamefont{Katzgraber et~al.}(2001)\citenamefont{Katzgraber,
  Palassini, and Young}}]{katzgraber:01}
\bibinfo{author}{\bibfnamefont{H.~G.} \bibnamefont{Katzgraber}},
  \bibinfo{author}{\bibfnamefont{M.}~\bibnamefont{Palassini}},
  \bibnamefont{and} \bibinfo{author}{\bibfnamefont{A.~P.} \bibnamefont{Young}},
  \emph{\bibinfo{title}{{M}onte {C}arlo simulations of spin glasses at low
  temperatures}}, \bibinfo{journal}{Phys. Rev. B}
  \textbf{\bibinfo{volume}{63}}, \bibinfo{pages}{184422}
  (\bibinfo{year}{2001}).

\bibitem[{\citenamefont{Katzgraber et~al.}(2009)\citenamefont{Katzgraber,
  Larson, and Young}}]{katzgraber:09b}
\bibinfo{author}{\bibfnamefont{H.~G.} \bibnamefont{Katzgraber}},
  \bibinfo{author}{\bibfnamefont{D.}~\bibnamefont{Larson}}, \bibnamefont{and}
  \bibinfo{author}{\bibfnamefont{A.~P.} \bibnamefont{Young}},
  \emph{\bibinfo{title}{Study of the de {A}lmeida-{T}houless line using
  power-law diluted one-dimensional {I}sing spin glasses}},
  \bibinfo{journal}{Phys. Rev. Lett.} \textbf{\bibinfo{volume}{102}},
  \bibinfo{pages}{177205} (\bibinfo{year}{2009}).

\bibitem[{\citenamefont{Dorogovtsev et~al.}(2008)\citenamefont{Dorogovtsev,
  Goltsev, and Mendes}}]{dorogovtsev:08}
\bibinfo{author}{\bibfnamefont{S.~N.} \bibnamefont{Dorogovtsev}},
  \bibinfo{author}{\bibfnamefont{A.~V.} \bibnamefont{Goltsev}},
  \bibnamefont{and} \bibinfo{author}{\bibfnamefont{J.~F.~F.}
  \bibnamefont{Mendes}}, \emph{\bibinfo{title}{{Critical phenomena in complex
  networks}}}, \bibinfo{journal}{Rev. Mod. Phys.}
  \textbf{\bibinfo{volume}{80}}, \bibinfo{pages}{1275} (\bibinfo{year}{2008}).

\bibitem[{com({\natexlab{e}})}]{comment:tc}
\bibinfo{note}{Note that all our numerical estimates for the critical
  temperatures are approximately 5\% above the analytically-determined values.
  Further studies would be needed to better understand this small deviation.
  The small systematic deviation could be, for example, attributed to
  corrections to scaling or the fact that the analytical results take all
  graphs to be equally weighted, whereas this is not the case in the numerical
  results. However, we emphasize that the agreement is remarkably good.}

\bibitem[{\citenamefont{Katzgraber et~al.}(2006)\citenamefont{Katzgraber,
  K\"orner, and Young}}]{katzgraber:06}
\bibinfo{author}{\bibfnamefont{H.~G.} \bibnamefont{Katzgraber}},
  \bibinfo{author}{\bibfnamefont{M.}~\bibnamefont{K\"orner}}, \bibnamefont{and}
  \bibinfo{author}{\bibfnamefont{A.~P.} \bibnamefont{Young}},
  \emph{\bibinfo{title}{{Universality in three-dimensional Ising spin glasses:
  A Monte Carlo study}}}, \bibinfo{journal}{Phys. Rev. B}
  \textbf{\bibinfo{volume}{73}}, \bibinfo{pages}{224432}
  (\bibinfo{year}{2006}).

\bibitem[{\citenamefont{Wittmann and Young}(2012)}]{wittmann:12}
\bibinfo{author}{\bibfnamefont{M.}~\bibnamefont{Wittmann}} \bibnamefont{and}
  \bibinfo{author}{\bibfnamefont{A.~P.} \bibnamefont{Young}},
  \emph{\bibinfo{title}{{{Spin glasses in the nonextensive regime}}}},
  \bibinfo{journal}{Phys. Rev. E} \textbf{\bibinfo{volume}{85}},
  \bibinfo{pages}{041104} (\bibinfo{year}{2012}).

\end{thebibliography}

\end{document}